\documentclass{jpp}
\usepackage{graphicx}
\usepackage{mathrsfs}
\usepackage{subfig}

\usepackage{bm} 
\usepackage[normalem]{ulem}

\paragraph{

\usepackage{url}

\usepackage[utf8]{inputenc}
\usepackage[T1]{fontenc}
\usepackage{amsmath}

\shorttitle{Analysis of Impurity Clustering in Tokamak Edge Plasma}
\shortauthor{Z. Lin, T. Maurel--Oujia, B. Kadoch, S. Benkadda and K. Schneider}

\title{Tessellation-based Analysis of Impurity Clustering in the Edge Plasma of Tokamaks}

\author{Z. Lin\aff{1},
 T. Maurel--Oujia\aff{1},
  B. Kadoch\aff{2} \\
 S. Benkadda\aff{3}
 \and K. Schneider\aff{1}\corresp{\email{kai.schneider@univ-amu.fr}}}

\affiliation{\aff{1}Aix-Marseille Universit\'e, CNRS, I2M, UMR 7373, 13453 Marseille, France
\aff{2} Aix-Marseille Universit\'e, CNRS, IUSTI, UMR 7343, 13453 Marseille, France
\aff{3} Aix-Marseille Universit\'e, CNRS, PIIM, UMR 7345, 13397 Marseille, France}

\begin{document}

\maketitle

\begin{abstract}
Confinement quality in fusion plasma is significantly influenced by the presence of heavy impurities, which can lead to radiative heat loss and reduced confinement. This study explores the clustering of heavy impurity, \textit{i.e.}, Tungsten in edge plasma, using high-resolution direct numerical simulations of the Hasegawa--Wakatani equations.    We use Stokes number to quantify the inertia of impurity particles. It is found that particle inertia will cause spatial intermittency in particle distribution and the formation of large-scale structures, \textit{i.e.}, the clustering of particles.  The degrees of clustering are influenced by Stokes number. To quantify these observations, we apply a modified Voronoi tessellation, which assigns specific volumes to impurity particles. By determining time changes of these volumes, we can calculate the impurity velocity divergence, which allows to assess the clustering dynamics. To quantify the clustering statistically,  several approaches are applied, such as probability density function (PDF) of impurity velocity divergence  and joint PDF of volume and divergence.\\

\end{abstract}
\section{Introduction}

Fusion reactors aim to achieve fusion reactions by confining and heating a plasma mixture of ions and electrons at extremely high temperatures.   Impurities in fusion plasmas can profoundly affect the performance and stability.  The impurities primarily originate from the interaction between the intense heat of the plasma and the tokamak's  walls, Their presence can lead to significant radiation losses and make it challenging to maintain the requisite temperatures for fusion reactions~\citep{Wesson2011}. These impurities can also dilute the primary fusion fuel, thereby reducing the overall fusion reaction rate. Additionally, strong radiation on the plasma's edge can drastically reduce electrical conductivity, leading to disruptions in the  current profile inside the $q=2$ surface~\citep{Stangeby2000}. Moreover, before reaching density limits, plasmas often develop MARFEs (multifaceted asymmetric radiation from the edge) which produce localized radiation, usually at the inner wall or the X-point, caused by strong impurity radiation near the edge~\citep{Stangeby2000}.

The behavior of plasma flow  within reactors, specifically in the edge region of a tokamak, is influenced by drift-wave turbulence and zonal flows. Our high-resolution simulations to understand this flow are based on the Hasegawa--Wakatani model  ~\citep{Hasegawa1983}, which provides insights into cross-field transport caused by electrostatic drift waves.  The electric field perpendicular to magnetic field lines, is particularly significant because it strongly drives cross-field fluxes, and influences edge pressure profiles and overall stability \citep{Zhang2020}. The Hasegawa--Wakatani equations are particularly useful for studying plasma turbulence, which is a major cause of energy loss in tokamak devices. They capture the essential physics of drift-wave turbulence. Despite their simplicity, the Hasegawa--Wakatani equations have proven to be a powerful tool for understanding the complex dynamics of plasma behavior in tokamak devices~\citep{horton2012turbulent}.

For the particles, the Lagrangian perspective has been of interest in recent years\citep{gheorghiu2024transport, Kadoch2022, Bos2010}. This perspective delves into transport properties by analyzing the trajectories of multiple tracer particles. Through numerical simulations, we solve equations about these particle trajectories within particular plasma flow velocity fields, like the $\bm{E} \times \bm{B}$ field. This method highlights the significant impact of coherent structures on transport. The interplay between eddy trapping and zonal shear flows results in non-diffusive transport~\citep{Krasheninnikov2008, delCastilloNegrete2004, vanMilligen2004, Krasheninnikov24a, Krasheninnikov24b}. Both experimental and numerical findings confirm the presence of these coherent structures in edge turbulence~\citep{Krasheninnikov2008}.

 In fusion plasma research, existing studies have focused on passive flow tracers in edge plasma, using the Hasegawa-Wakatani model, without taking inertial effects into account ~\citep{ Futatani2008b, Futatani2008a, Futatani2009a}.  In the work by ~\citet{Priego2005}, inertial effect is considered and a  fluid model for impurities is used, wherein the impurity particle   velocity is considered to be the sum of the $\bm{E} \times \bm{B}$  and polarization drifts.  The polarization drift, accounting for impurity particle inertia, is a higher-order correction to the $\bm{E} \times \bm{B}$ drift velocity and introduces compressible effects.   In contrast to fluid model,  we track   each impurity particle individually.  In addition,  we assume that  heavy impurity particles may “lag behind”  plasma flow velocity due to the significant inertia.  Indeed, the role of inertia in particle advection by turbulent flows is  well-documented  in fluid dynamics, known to result in clustering around vortical structures ~\citep{Provenzale99}. Our study aims to analyze possible  clustering effects  for impurities in magnetized plasma.

A key element in understanding these heavy impurities is their tendency for self-organization. This self-organization is seen in clusters of impurities and void regions~\citep{Monchaux2910, Lin2024}. It can be mathematically quantified by deviation from Poissonian statistics~\citep{Oujia2020, MaurelOujia2023}. The goal is to investigate the statistical signature of clustering and void regions. Recent results for hydrodynamics turbulence are promising ~\citep{Matsuda2021}. Finite-time measures to quantify divergence and the rotation of the particle velocity by determining respectively the volume change rate of the Voronoi cells and their rotation were proposed in~\citet{Oujia2020, MaurelOujia2023} and applied in the context of hydrodynamic turbulence. Here we apply them to characterize the dynamics of the self-organization of heavy impurities in plasma  and assess their clustering and void formation.

The paper is structured as follows: Sec.~\ref{sec: Models and Method} outlines the theoretical basis used for simulation, tessellation-based analysis method and numerical setup. Sec.~\ref{sec: Results}  details our numerical simulation results and tessellation-based analysis results. Finally, Sec.~\ref{sec: Conclusions} summarizes our findings and suggests potential directions for future investigations.

\section{\label{sec: Models and Method}Models for simulation and analysis method}
\subsection{Hasegawa--Wakatani model}

\begin{figure} 
 \centerline{\includegraphics[width=1.0\textwidth]{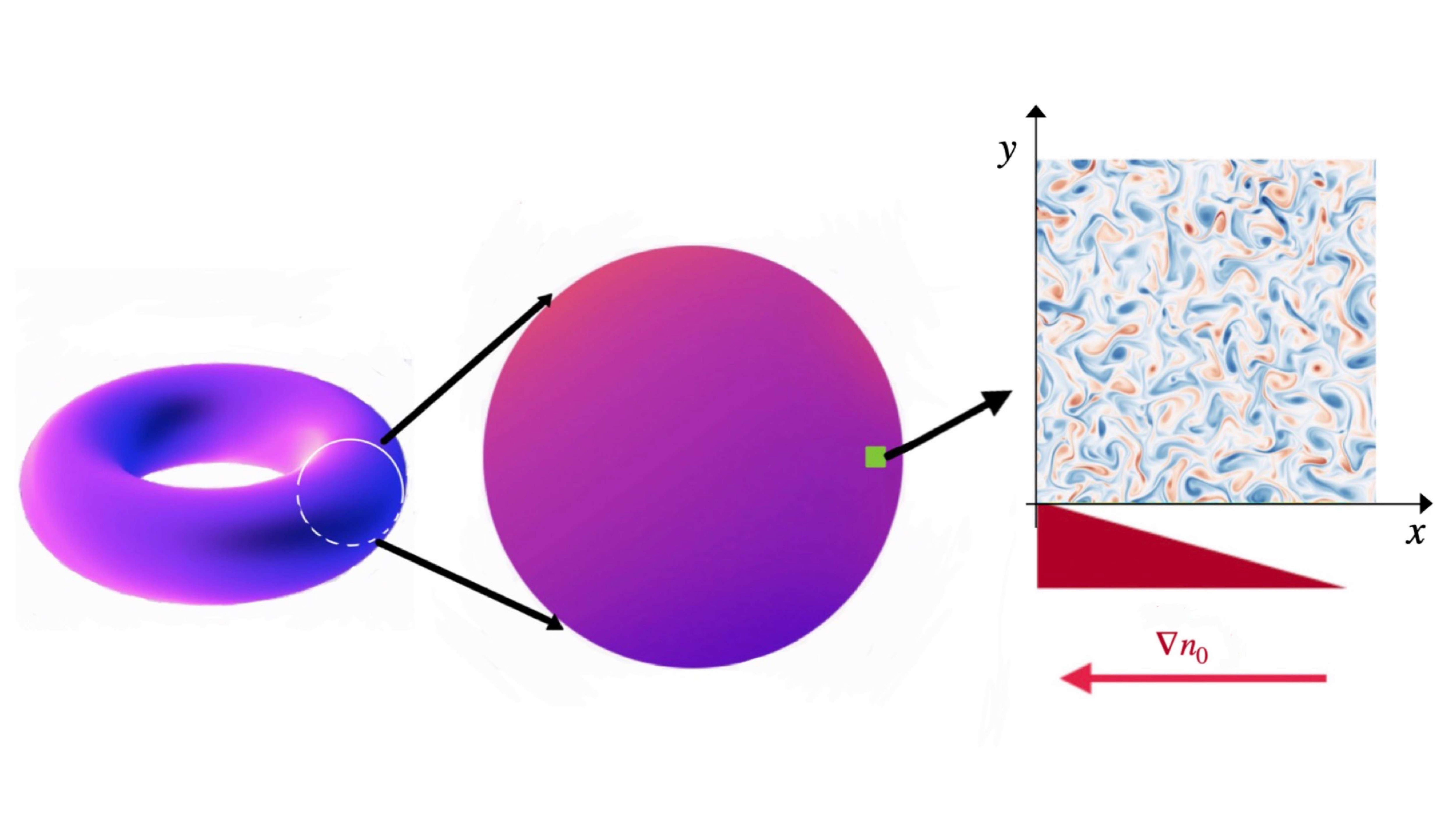}}
        \caption[]{Illustration of the two-dimensional slab geometry. In the tokamak edge, the 2D slab geometry flow configuration is depicted using the Hasegawa–-Wakatani system. Here, the radial direction is represented by $x$, while $y$ is the poloidal direction. There is an imposed mean plasma density gradient $\nabla{n_0}$ in the radial direction. The 2D flow is computed within a square domain measuring 64 Larmor radii ($\rho_s$) on each side, as indicated by the green outline. On the right, a vorticity field is displayed.}
        \label{fig:slab geometry}
 \end{figure}

 The model for our numerical simulations is based on Hasegawa--Wakatani (HW) equations for plasma edge turbulence driven by drift-wave instability~\citep{Hasegawa1983}.  Our focus in this study is on the two-dimensional slab geometry of the HW model, see {\it e.g.} \citet{Kadoch2022}. The 2D HW model is a representative paradigm for understanding the nonlinear dynamics of drift-wave turbulence.  Fig.~\ref{fig:slab geometry} depicts a representation of the flow configuration.

The magnetic field lines are assumed to be straight and perpendicular to the slab. Ions are considered cold, ignoring temperature gradient effects.  The key variables are normalized as detailed in the work done by~\citet{Bos2010} and~\citet{Futatani2011},
$$
x/\rho_{\mathrm{s}} \rightarrow  x,\quad  \omega_{\mathrm{ci}} t \rightarrow t,  \quad e \phi / T_{\mathrm{e}} \rightarrow \phi,\quad n_{1} / n_{0} \rightarrow n
$$   $\rho_s$ is the ion Larmor radius at electron temperature $T_{\mathrm{e}}$. It is  defined as $\rho_s = \frac{\sqrt{T_e/m_i}}{\omega_{ci}}$, where $\omega_{ci}$ is the ion cyclotron frequency.  Here $n_{1}$ and $n_{0}$ represent plasma density fluctuation and equilibrium density, while $\phi$ indicates electrostatic potential fluctuation. The HW model consists of two equations that describe the  evolution of plasma potential and fluctuating plasma density,  respectively:

\begin{equation}
\left(\frac{\partial}{\partial t}- \mu_{\nu} \nabla^{2}\right) \nabla^{2} \phi=\left[\nabla^{2} \phi, \phi\right]+c(\phi-n),
\label{eq:phi}
\end{equation}

\begin{equation}
\left(\frac{\partial}{\partial t}-\mu_{D} \nabla^{2}\right) n=[n, \phi]-\kappa \frac{\partial \phi}{\partial y}+c(\phi-n),
\label{eq:n}
\end{equation} 
where $\mu_{D}$ is cross-field diffusion coefficient and $\mu_{\nu}$ is kinematic viscosity.
The term $\kappa$ defined as $\kappa \equiv - \partial_{x} \ln(n_{0})$, is a measure of the plasma  density gradient. Here $\kappa$ is assumed constant,  implying $n_{0} \propto \exp (-\kappa x)$. 
The Poisson bracket is defined 
as: $[A, B]=\frac{\partial A}{\partial x} \frac{\partial B}{\partial y}-\frac{\partial A}{\partial y} \frac{\partial B}{\partial x}$. In these equations, the electrostatic potential $\phi$ is  the stream-function for the $\bm{E}  \times \bm{B}$ velocity, represented by $\bm{u}=\nabla^{\perp} \phi$, where $\nabla^{\perp}=\left(-\frac{\partial}{\partial y}, \frac{\partial}{\partial x}\right)$. So, $u_x=-\frac{\partial \phi}{\partial y} $ and $u_y=\frac{\partial \phi} {\partial x}$ with vorticity $\omega=\nabla^2 \phi$.

In the equations, $c$ is so-called adiabaticity parameter which measures the parallel electron response. It is defined as:

\begin{equation}
c=\frac{T_{e} k_{\parallel}^{2}}{\mathrm{e}^{2} n_{0} \eta \omega_{c i}}
\end{equation}

With $\eta$ as electron resistivity and $k_{\parallel}$ as the effective parallel wavenumber, parameter $c$ determines the phase difference between electrostatic potential and plasma density fluctuations. The model described above is the classical Hasegawa--Wakatani model (cHW). For $c\gg 1$ (adiabatic limit), the model reduces to the Hasegawa--Mima equation, where electrons follow a Boltzmann distribution. At $c \ll 1$  (hydrodynamic limit), the system reduces to a form that is analogous to two-dimensional Navier--Stokes equation, where density fluctuations are passively influenced by the $\bm{E} \times \bm{B}$ drift. Notably, for $c \sim 1$ (quasi-adiabatic regime), a phase shift between the potentials and densities is observed, enabling particle transport and reflecting a complex interplay between them. 
To obtain zonal flow, a revised version of the model, known as the modified Hasegawa--Wakatani model (mHW), can be considered. This modification was introduced in ~\citet{Pushkarev2013} which consists in setting the coupling term $c(\phi -n)$ to zero for modes $k_y= 0$. In this paper, we will focus on the quasi-adiabatic regime, \textit{i.e.}, $c = 0.7$ (cHW),  which is relevant to the edge plasma of tokamaks~\citep{Bos2010}. For readers interested in the results of other regimes, we refer to Appendix~\ref{appendix: different regimes}.

\subsection{Impurity particles model}

Earlier research ~\citep{Futatani2008b,Futatani2008a,  Futatani2009a, Kadoch2010, Priego2005}  assumed that impurity particles, due to no inertia, act as "passive tracers" aligning closely with fluid flow. Yet, for heavier particles, this assumption might not always hold, particularly when the particles have significant mass, leading to noticeable inertial effects. To address this, the Stokes number ($St$) can be introduced, see e.g. \citet{Oujia2020}, a dimensionless parameter that quantifies particle inertia in fluid flow. The Stokes number is defined as:

\begin{equation}
St = \frac{\tau_{p}}{\tau_{\eta}},
\end{equation}
where $\tau_{p}$ denotes the impurity particle's relaxation time, it measures the time the particle takes to adapt to fluid flow alterations.  A smaller $\tau_{p}$ means the particle adapts faster. On the other hand, $\tau_{\eta}$ is the characteristic timescale for turbulence, indicating how fast the fluid flow changes. A smaller $\tau_{\eta}$ suggests more rapid turbulence changes. A high Stokes number means that a particle's movement is primarily driven by inertia, causing it to keep its direction despite fluid changes. In previous studies~\citep{Futatani2008b, Futatani2008a, Futatani2009a},  the impurity particles are treated as passive tracers of the flow, indicating $\tau_p = 0$.  However, here we consider heavy impurity particles with $\tau_p > 0$, for instance Tungsten. 

The heavy impurity particles are modeled as charged point particles. They are considered as test particles, which means that there is no impurity-impurity interaction and impurities have no impact on the plasma dynamics,  while the plasma flow velocity affects the motion of impurities (one-way coupling).  Impurities in our model are assumed to be cold, just as  the treatment of  ions in bulk plasma. Thus thermal motion of impurities  could be neglected for simplicity. The impurity particles experience not only the Lorentz force due to the electric and magnetic fields but also  drag force resulting from momentum transfer with plasma ions. Positive plasma ions transfer momentum to heavy charged impurity particles by interacting with electrostatic potential in the vicinity of the impurity particle. Electrons transferring momentum to impurities is negligible due to their small mass.  Coulomb scattering theory  provides, macroscopically, an approximation of a drag force $F_{drag}$  dependent on the relative velocity  between the macroscopic velocity of the plasma ion and the velocity of the impurity particle~\citep{Kilgore93}: $F_{drag} =  K_{\mathrm{mt}} n_i m_i v_r$,  where $v_r$ is the relative velocity  between the macroscopic velocity of the plasma ions and the impurity particles, $K_{\mathrm{mt}}$ is the momentum transfer coefficient, $n_i$ is the plasma ion density, and $m_i$ is the ion mass of the plasma.  This  drag force tends to reduce the velocity difference, causing the impurity particles to relax to the plasma flow velocity.  The relaxation time  of the impurity particle  $\tau_p$,  is  the period during which impurity particles adjust to the plasma flow due to the drag force.  In our model, we approximate this drag force as  $m_p(u_p-v_p)/\tau_p$ , which is  proportional to the relative velocity between  the plasma flow and impurities.  The relaxation time can be expressed as: 
$ 
\tau_p=\frac{m_p}{K_{\mathrm{mt}} n_i m_i }.
$ 
 This   relation highlights that $\tau_p$ is inversely proportional to the plasma ion density and momentum transfer coefficient,  while directly proportional to the impurity mass. Given this relationship, as the ion density $n_i$ increases, the relaxation time decreases, leading to faster coupling of the impurity particles to the plasma flow. Conversely, larger impurity masses result in longer relaxation times, reflecting the slower response of heavier particles to the plasma flow.  According to Newton's second law:

\begin{equation}
m_{p} \frac{d \bm{v}_{p,j}}{dt} = \frac{m_{p}(\bm{u}_{p,j}- \bm{v}_{p,j})  }{\tau_p} + Ze(\bm{E}(\bm{x}_{p,j}) + \bm{v}_{p,j} \times \bm{B})
\label{eq: equation of motion 1}
\end{equation}
\begin{equation}
\bm{E}(\bm{x}_{p,j})= -\nabla \phi(\bm{x}_{p,j}) 
\label{eq: electric field}
\end{equation}
Equation \ref{eq: equation of motion 1} describes the forces exerted on each heavy impurity particle, denoted by $j$ ($j=1, \ldots, N_p$, with $N_p$ being the total number of particles). $Z$  is charge state of the impurity particle and $e$ is the elementary charge. The variable $\bm{v}_{p,j}$ is the $j$-th particle's velocity, while $\bm{u}_{p,j}$, $\bm{E}(\bm{x}_{p,j})$ and $\nabla \phi(\bm{x}_{p,j})$ representing the fluid velocity,  Electric field and gradient of potential at the particle's location $\bm{x}_{p,j}$, respectively.   Let us note that when $\tau_p = 0$, we recover the situation studied in previous works~\citep{Futatani2008b, Futatani2008a, Futatani2009a}. In this case, the impurity particles are considered as passive tracers of the flow, \textit{i.e.} $\bm{v}_{p,j} \equiv \bm{u}_{p,j}$.  Applying the same normalization as those used for the HW equation, we obtain:
\begin{equation}
\frac{d \bm{v}_{p,j}}{d t}=\frac{(\bm{u}_{p,j}-\bm{v}_{p,j})}{\tau_p}+ \ \alpha (-\nabla \phi(\bm{x}_{p,j})+\bm{v}_{p,j} \times \bm{b})
\label{eq: equation of motion 2}
\end{equation}
where $\alpha=\frac{Zm_i}{m_{p}}$, $m_i$ is the ion mass of the plasma, and $\bm{b}$ represents the unit vector along the direction of the magnetic field. Combining with the equation:
\begin{equation}
\frac{d \bm{x}_{p,j}}{dt} = \bm{v}_{p,j}  
\label{eq: equation of motion 3}
\end{equation}
it is thus possible to compute the trajectory of the impurity particles.

\subsection{Modified Voronoi tessellation}
 
\begin{figure}
\centering
\includegraphics[width=1.0\textwidth]{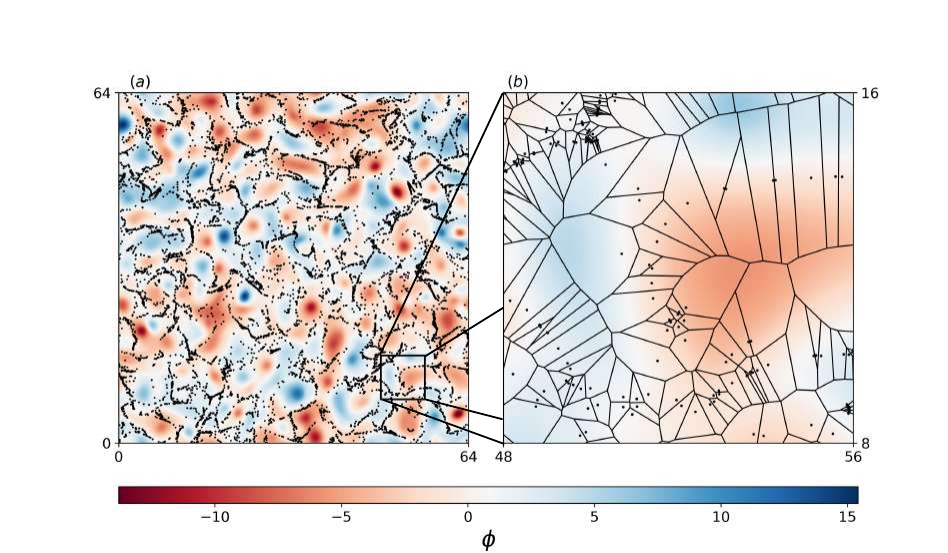}\caption{(a) Electric potential $\phi$ (stream function) and $10^4$ superimposed impurity particles for $St = 1$ in the case $c = 0.7$ (cHW). (b) A magnified view with modified  Voronoi tessellation. Large cells denote void regions, small cells represent clusters. }
\label{tessellation}
\end{figure}

Impurity particles satisfy the conservation equation:
 
\begin{equation}
D_t n_p=-n_p \nabla \cdot {\bm v_p},
\label{continuity}
\end{equation}
where  $D_t$ is the Lagrangian derivative,  $n_p$ is the impurity particle number density  and $\bm v_p$ the particle velocity.  From this equation, the divergence of particle velocity $\nabla \cdot {\bm v_p} = -\frac{D_t n}{n}$  is crucial to understand. A positive divergence ($\nabla \cdot {\bm v_p} > 0 $)  leads to density decrease, whereas a negative divergence,\textit{i.e.} convergence  ($\nabla \cdot {\bm v_p} < 0 $) causes an increase in density.  To quantify the divergence, we employ the modified version of the classical Voronoi tessellation technique, which uses the center of gravity, instead of the circumcenter of the Delaunay cell to define  Voronoi cell vertices.  The purpose of this method is to assign a volume to each impurity particle, enabling the subsequent calculation of the divergence.   More details about the classical and modified Voronoi tessellation and their difference can be found in \citet{Oujia2020, MaurelOujia2023}.
The modified Voronoi tessellation divides a plane into regions according to the positions of impurity particles,  which allows to assign a cell, called "modified Voronoi cell",  and thus a corresponding volume $V$, to each particle. Each cell can be considered as the region of influence of an impurity particle. 
In our two-dimensional scenario, the volume of the cell is determined by calculating its area. Figure \ref{tessellation} shows the modified Voronoi tessellation. In the figure, clustered particles are represented by small cells and void regions are represented by large cells.

To compute the divergence of impurity particle velocity ${\cal D}$
we analyze particle distributions at two time instants $t^k$ and $t^{k+1}=t^k+\Delta t$, where $\Delta t$ is the time step  of the simulation and $k$ is the discrete time index. Each particle distribution snapshot is assessed using modified Voronoi tessellation.
The inverse of volume, $1/V$, approximates the local particle number density $n_{p}$ in the discrete setting. It is proved  that ~\citep{Oujia2020, MaurelOujia2023}:

\begin{equation}
{\cal D} ({\bm v_p}) = -\frac{1}{n_{p}} D_t n_{p} =  \frac{2}{\Delta t} \frac{V^{k+1}-V^k}{V^{k+1}+V^k} + O\left(\frac{1}{N_{p}}, \Delta t\right) 
\label{divergence}
\end{equation}

 Thus by measuring the volume change of each particle between $t^k$ and $t^{k+1}$, we can determine the discreet divergence of its velocity ${\cal D}({\bm v_p}$). The velocity divergence of impurity particles measures the rate of volume change for a particle group over time, indicating how much the particle velocity field is spreading out or converging at a particular point in space.

\subsection{\label{sec: Numerical Simulations}Numerical setup}

The simulation was conducted within a domain that spans an area $A$ of $64 \times 64$ with periodic boundary condition. The domain was discretized to a resolution $R$ of  $1024 \times 1024$ grid points. The time step $\Delta t$ was $5\times10^{-4}$. The number of impurity particles $N_{p}$ was $10^6$. The adiabaticity parameter $c$ is $0.7$.  The values of specific physical parameters are listed in Table \ref{tab: flow simulation parameters}. The characteristic time scale of turbulence, $\tau_{\eta}$,  is defined as $1 / \sqrt{2 Z_{\rm m s}}$, with $Z_{\rm m s}$ denoting one half of the mean-square vorticity.  In the simulation,  $\tau_{\eta} = 0.35$  in the statistically steady flow.  The simulations are based on the work of ~\citet{Kadoch2022} where the flow is initialized with Gaussian random fields. Here we start the simulations with a flow already being in the statistically steady state (\textit{i.e.} after the long transition phase of drift-wave instabilities, see \citet{Kadoch2022}) and one million uniformly distributed heavy impurity particles with random velocity are injected.

For the impurity particles, we will focus on Tungsten $_{74}^{184}\mathrm{W}^{20+}$, a heavy one that can be made from the material that is exposed to the high heat flux in  ASDEX-Upgrade~\citep{krieger1999conclusions}  and  EAST~\citep{yao2015design}, expecting the same situation for ITER~\citep{pitts2009status}.  For $_{74}^{184}\mathrm{W}^{20+}$, $\alpha = Zm_i/m_{p} = 0.22$. As for Stokes number $St =  \tau_p/\tau_\eta$, $\tau_\eta = 0.35$ in the statistically steady state of of flow for $c = 0.7$ (cHW),  but the exact value of $\tau_p$ for $_{74}^{184}\mathrm{W}^{20+}$ is unknown. To resolve this, we explore a range of $\tau_p$ values spanning several orders of magnitude, resulting in Stokes numbers: $St = 0, 0.05, 0.5, 1, 5, 10 $ and $50$. This parametric approach allows us to systematically explore the dynamics of Tungsten impurity, from particles that are tightly coupled to the flow ($St \ll 1$) to those that are essentially independent of the flow ($St \gg 1$). The simulation parameters are listed in  Table~\ref{tab: stokes number}.

\begin{table}
  \begin{center}
  \setlength{\tabcolsep}{8pt} 
    \begin{tabular}{cccccccc}
      A & R & $\Delta t$ & $N_{p}$ &  $\mu_{D}$ &   $\mu_{\nu}$ & $\kappa$ & $c$\\[3pt]
      $64\times 64$ & $1024 \times 1024$ & $5 \times 10^{-4}$ &  $10^{6}$ &  $5\times10^{-3}$ & $5\times10^{-3}$ & 1  & 0.7\\
    \end{tabular}
    \caption{Simulation parameters for the flow. $A$: Domain area; $R$: Grid resolution; $\Delta t$: Time step; $N_{p}$: Number of impurity particles; $\mu_{D}$: Diffusion coefficient;  $\mu_{\nu}$: Kinematic viscosity; $\kappa \equiv - \partial_{x} \ln(n_{0})$,
    is a measure of the plasma density gradient; $c$: Adiabaticity parameter}
    \label{tab: flow simulation parameters}
  \end{center}
\end{table}

\begin{table}
  \begin{center}
  \setlength{\tabcolsep}{12pt} %
    \begin{tabular}{ccccccccc}
      $St$& 0 & 0.05 & 0.25 & 0.5 &  1 &  2 & 5 & 50  
    \end{tabular}
    \caption{Stokes number used in the simulation }
    \label{tab: stokes number}
  \end{center}
\end{table}

The system of HW equations, specifically Eq.~\eqref{eq:phi} and Eq.~\eqref{eq:n} are solved using a classical pseudospectral method~\citep{canuto2007spectral}. The equations governing the motion of the impurity particles, namely Eq.~\eqref{eq: equation of motion 2} and Eq.~\eqref{eq: equation of motion 3}, are solved using a second-order Runge--Kutta (RK2) scheme and linear interpolation is used for computing the fluid velocity at the particle position. The code of this study builds upon that used in the research done by ~\citet{Kadoch2022}.

\section{\label{sec: Results}Results}
\subsection{Simulation results}
\begin{figure}
\centering
\includegraphics[width=0.5\textwidth]{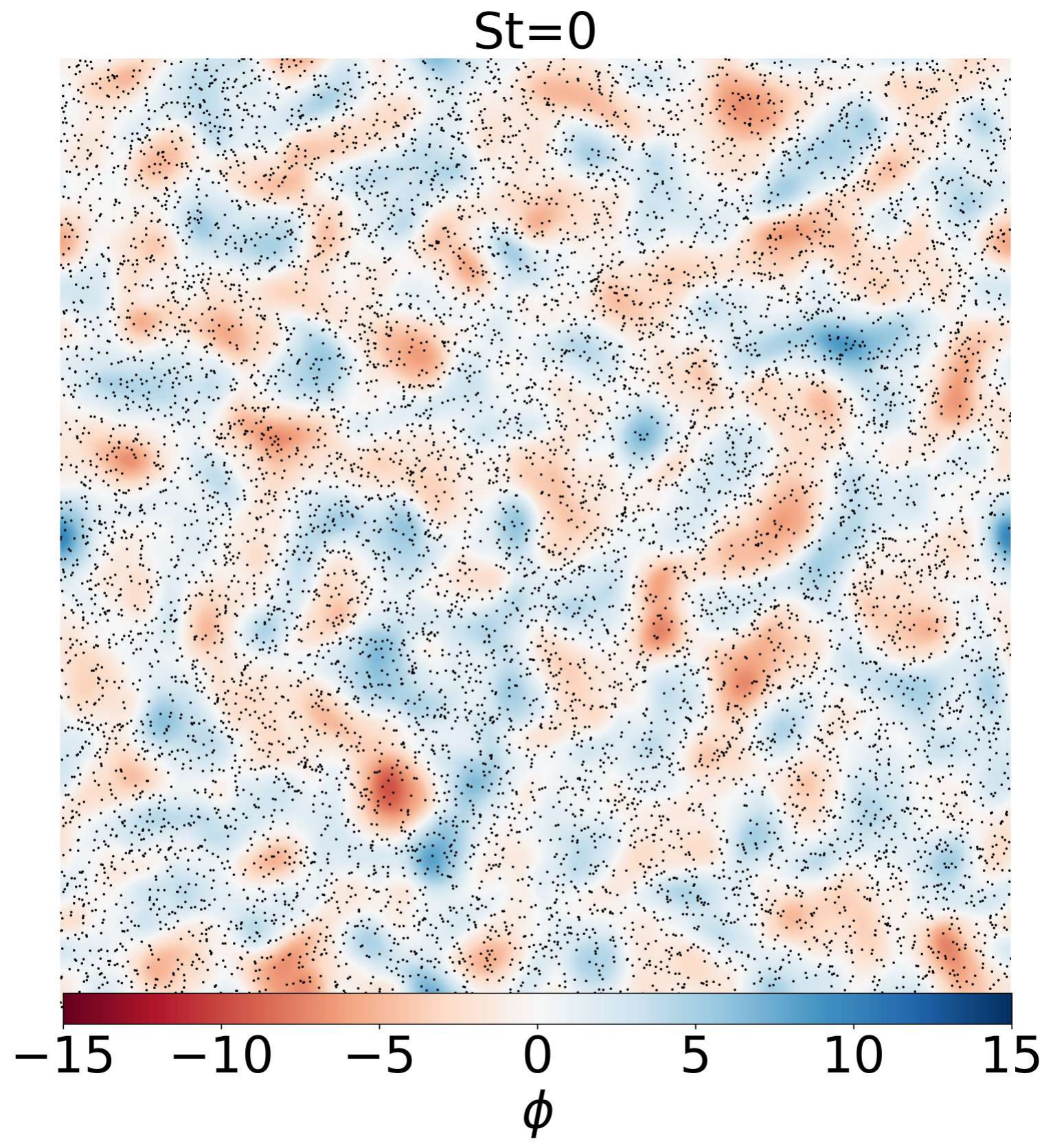}
\caption{ Electric potential fields $\phi$ (stream function) superimposing $10^4$ impurity particles (out of
$10^6$) for $St = 0$  in statistically steady state within the quasi-adiabatic regime ($c = 0.7$, cHW).}
\label{fig: vorticity and particles:St = 0}
\end{figure}

\begin{figure}
        \centering
        \includegraphics[width=0.9\linewidth]{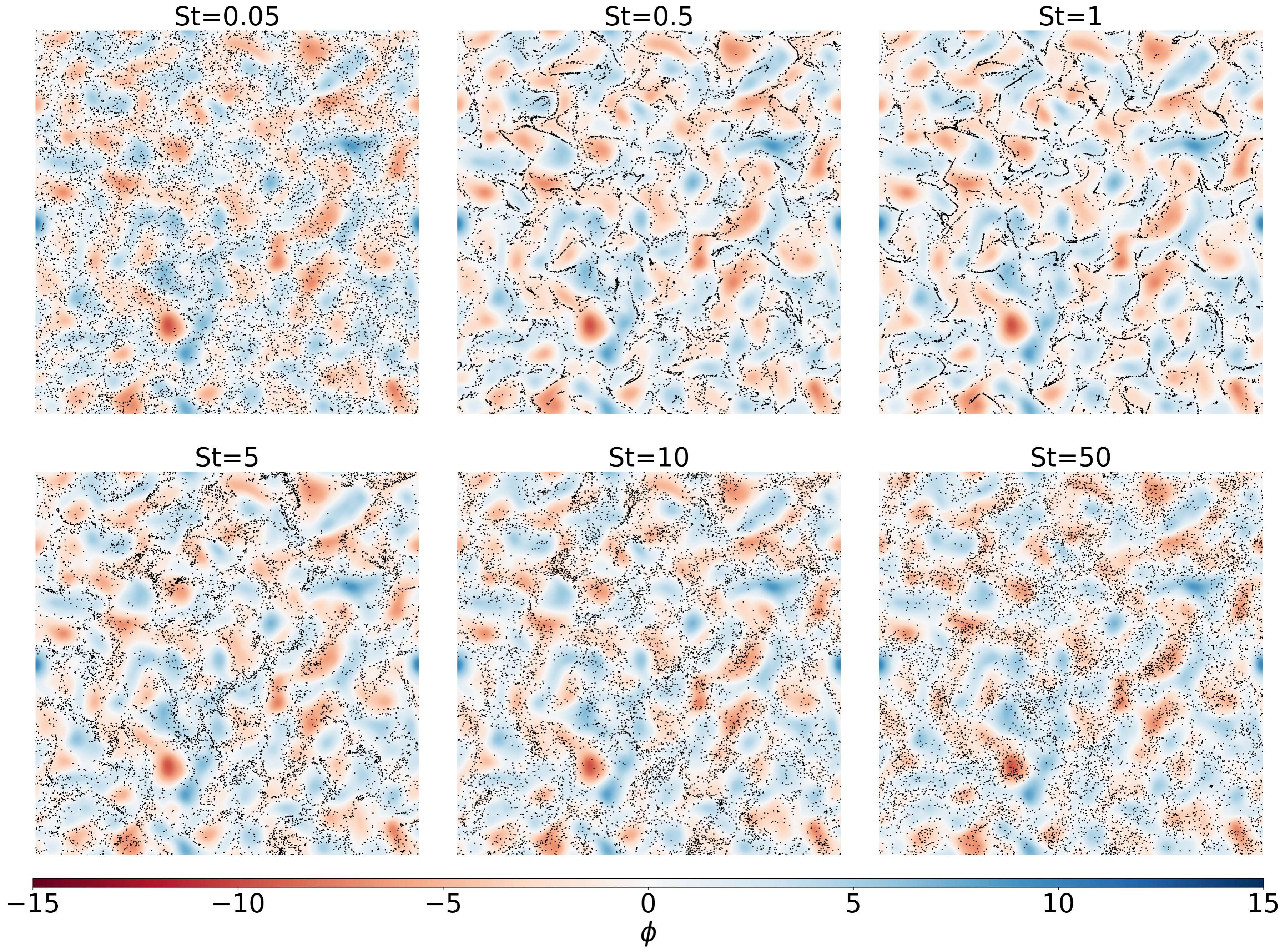}
        \caption[]{Electric potential fields $\phi$ (stream function) superimposing $10^4$ impurity particles (out of
$10^6$) for various Stokes numbers in statistically steady state within the quasi-adiabatic regime ($c = 0.7$, cHW).}
        \label{fig: vorticity and particles: c =0.7}
 \end{figure}

The analyzed data set consists of the electric potential $\phi$ (stream function), particle position and velocity data generated by direct numerical simulation (DNS). The contour curves of constant $\phi$ represent the streamlines of the flow field and provide a clear visualization of fluid flow in two dimensions. Closed contour curves indicate the presence of vortices. Figure~\ref{fig: vorticity and particles:St = 0} illustrates the $\phi$ fields and the behavior of impurity particles in  statistically steady state within the quasi-adiabatic regime ($c = 0.7$, cHW), where impurity particles are considered as passive tracers  without considering inertial effect ($St = 0$).  As shown by Figure~\ref{fig: vorticity and particles:St = 0}  impurity particles are distributed randomly.   Figure~\ref{fig: vorticity and particles: c =0.7} shows that with increasing Stokes numbers ($St =  0.05,  0.5, 1 $), particle inertia begins to dominate, causing them to deviate from the fluid streamlines due to Coriolis force induced by the vortices. This deviation causes them to concentrate where the vorticity is minimal, typically at vortex peripheries.  For even higher Stokes numbers ($St = 5, 10, 50$), the impurity particles tend to maintain their trajectories, moving more  ballistically  and exhibiting reduced clustering. From Eqn.~\eqref{eq: equation of motion 2}, when the coefficient of the flow effect term, $ 1/\tau_p$ =  1/($ \tau_\eta St)$, is much larger than the coefficient term of the Lorentz force $\alpha$,  the particle dynamics are primarily governed by the flow, while the Lorentz force plays a relatively minor role. For $St = 0.05,  0.5, 1$,    $ 1/\tau_p = 1/(0.35 St ) \gg \alpha = 0.22 $,  the particles are more strongly influenced by the fluid flow than by the electromagnetic forces. However, as the Stokes number increases to values such as $St = 10, 50$,  the particles are less coupled to the fluid flow  and the electromagnetic forces start to play an important role. As shown in  Figure~\ref{fig: vorticity and particles: c =0.7}, for case $St = 10$ and $St = 50$, the positively charged impurity particles are mostly located in negative potential regions (red region), as expected. The effects of different $\alpha$ values  are further explored in Appendix~\ref{appendix: Analysis}. 
 
  Let us note that in regions of impurity clustering, where the impurity density may locally increase, there is a risk that   clustering could lead to deviations from the assumption of quasi-neutrality  that is underlying the derivations of the HW model  ~\citep{Naulin06}.  This strict condition, which requires that the impurity density remains smaller than the bulk current divergence on the characteristic drift-wave time scale, cannot always be guaranteed in regions of intense clustering.   Nevertheless, as noted in ~\citet{Naulin06}, considering impurities   as test particles  is still valuable for understanding tendencies in impurity transport within a given flow field, and the clustering effects observed in our simulations provide   insights into impurity dynamics. We further acknowledge that resolving this discrepancy would require a more sophisticated model that accounts for the back-reaction of impurity densities on the plasma fields and their contribution to quasi-neutrality. Such a model would need to self-consistently address the plasma profile problem in the presence of impurities, which is beyond the scope of the present work, but remains a valuable direction for future research.

\subsection{Tessellation-based analysis results}
\begin{figure}
    \centering
    \subfloat[]{\includegraphics[width=0.48\textwidth]{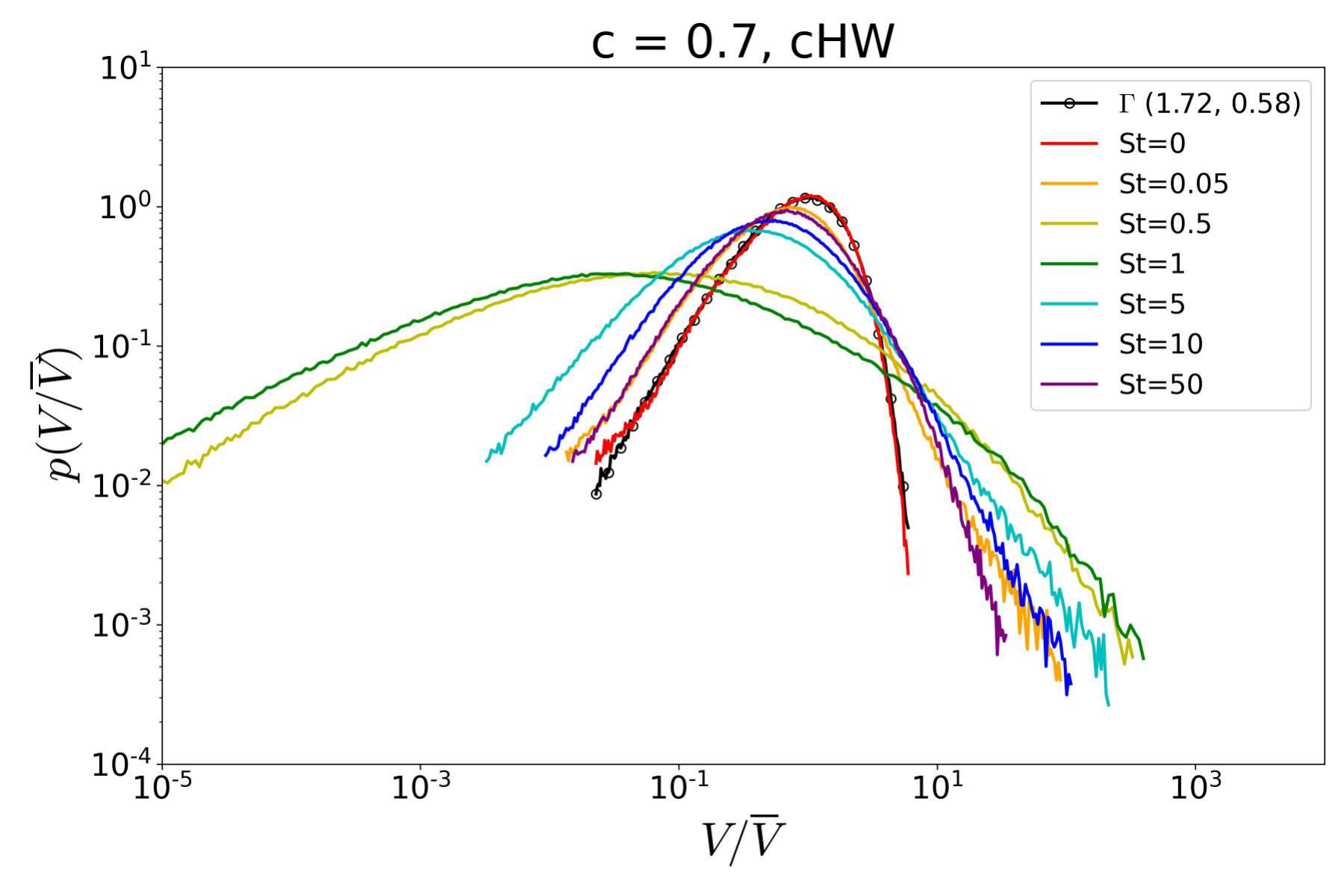}}\quad
    \subfloat[]{\includegraphics[width=0.48\textwidth]{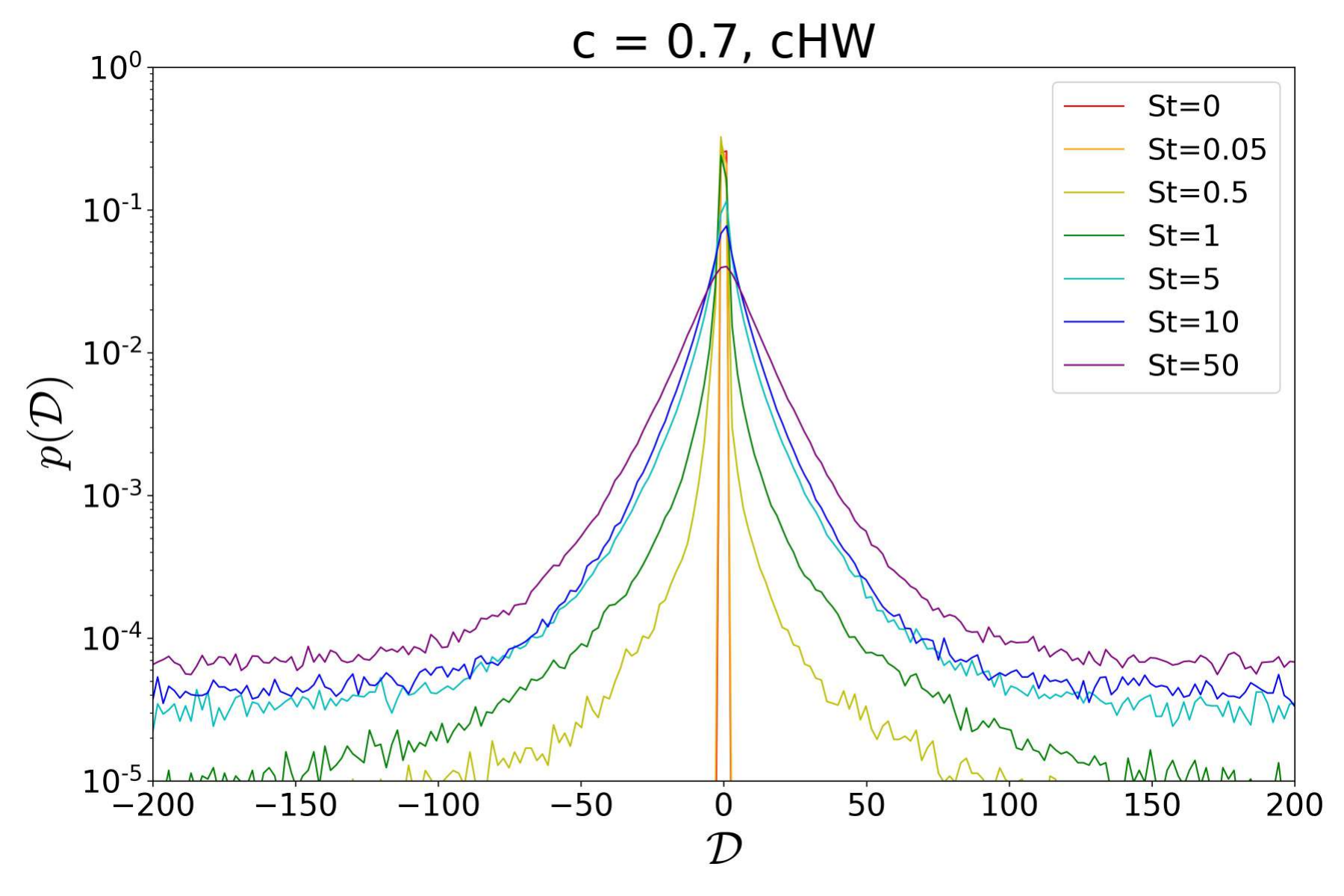}\label{fig: divergence_pdf}}
    \caption{(a) PDF of  volume normalized by the mean for different Stokes numbers against the gamma distribution. (b) PDF of the divergence of impurity particle velocity for different Stokes numbers for $_{74}^{184}\mathrm{W}^{20+}(\alpha = 0.22)$ in quasi-adiabatic regime ($c = 0.7$, cHW).}

    \label{fig: volume_divergence_pdf}
\end{figure}

We calculate the volume $V$  for each impurity particle, and then plot the Probability Density Function (PDF) of $V/\overline{V}$,  where $\overline{V}$ is the average volume, calculated as   $ \overline{V} = 64^2/N_{p}$ ($64^2$ is the domain area, $N_{p}$ is the total number of particles). Figure \ref{fig: volume_divergence_pdf}(a) shows the PDF of $V/\overline{V}$.

A random variable $X$ that is gamma-distributed with shape $k$ and scale parameter $\theta$ is denoted $X \sim \Gamma(k, \theta)$. The corresponding PDF is $f(x)=\Gamma(k)^{-1} \theta^{-k} x^{k-1} \exp (-x / \theta)$.
For particles that are randomly distributed, the PDF of the  volumes  follows a gamma distribution \citep{FerencNeda2007}.  For $St = 0$, using the maximum likelihood PDF estimation for fitting,  we find that the PDF of the  volumes  aligns well with the gamma distribution, specifically we have $\Gamma(1.72, 0.58)$. This suggests, as expected, that the impurity particles are randomly distributed when $St = 0$, indicating no clustering  of particles.  Let us note that the PDF of the volumes for $St = 0$ obtained using  classical Voronoi tessellation follows a gamma distribution $\Gamma(7/2, 2/7)$ \citep{FerencNeda2007}. Since we use a different method, the modified Voronoi tesselation, the fitting gives different values, \textit{i.e.}  $\Gamma(1.72, 0.58)$. The PDF of the   volume is typically used to classify "cluster cells" and "void cells" ~\citep{Monchaux2910}. A  cell below a certain threshold is considered a cluster cell, whereas a cell above this threshold is categorized as a void cell. In our study, the threshold is $V/\overline{V} \sim 0.5$. 
As shown by Figure~\ref{fig: volume_divergence_pdf}(a), when  $St$ increases, we observe an  increase in the number of cluster cells (with $V/\overline{V} < 0.5$)  However, once $St$ exceeds $St = 1$,  the number of  cluster  cells begins to decrease. This is also highlighted by Figure~\ref{fig: vorticity and particles: c =0.7} which shows that the particles are densely packed for $St =1$.

The PDF of the divergence of impurity particle velocity is displayed in figure \ref{fig: volume_divergence_pdf}(b). As shown in figure \ref{fig: volume_divergence_pdf}(b),  the divergence of the fluid velocity (represented by the divergence of the impurity particle velocity with $St = 0$) is  small but not zero. However, as we know,   in an incompressible continuous fluid, the divergence of the fluid velocity should be zero. The reason for this discrepancy lies in our computational method, which segments the fluid into separate pieces known as "modified Voronoi cells" as mentioned before. The deformation of a modified   Voronoi cell does not correspond to the deformation of a fluid volume in a continuous context. The figure implies that the divergence is attributable to   discretization errors. A negative divergence in a particle velocity field indicates that particles converge towards a point, thus increasing  density locally. Conversely, positive divergence shows that particles spread out from a point, reducing the local density. As shown in figure \ref{fig: volume_divergence_pdf}(b), for increasing Stokes number, the PDF of divergence ${\cal D}$ widens, reflecting stronger particles convergence and divergence activities.  The symmetry observed in Figure~\ref{fig: volume_divergence_pdf}(b) suggests that positive and negative divergence values are comparable. This balance is due to the conservation of impurity particles.

\begin{figure}
        \centering
        \includegraphics[width=1.\linewidth]{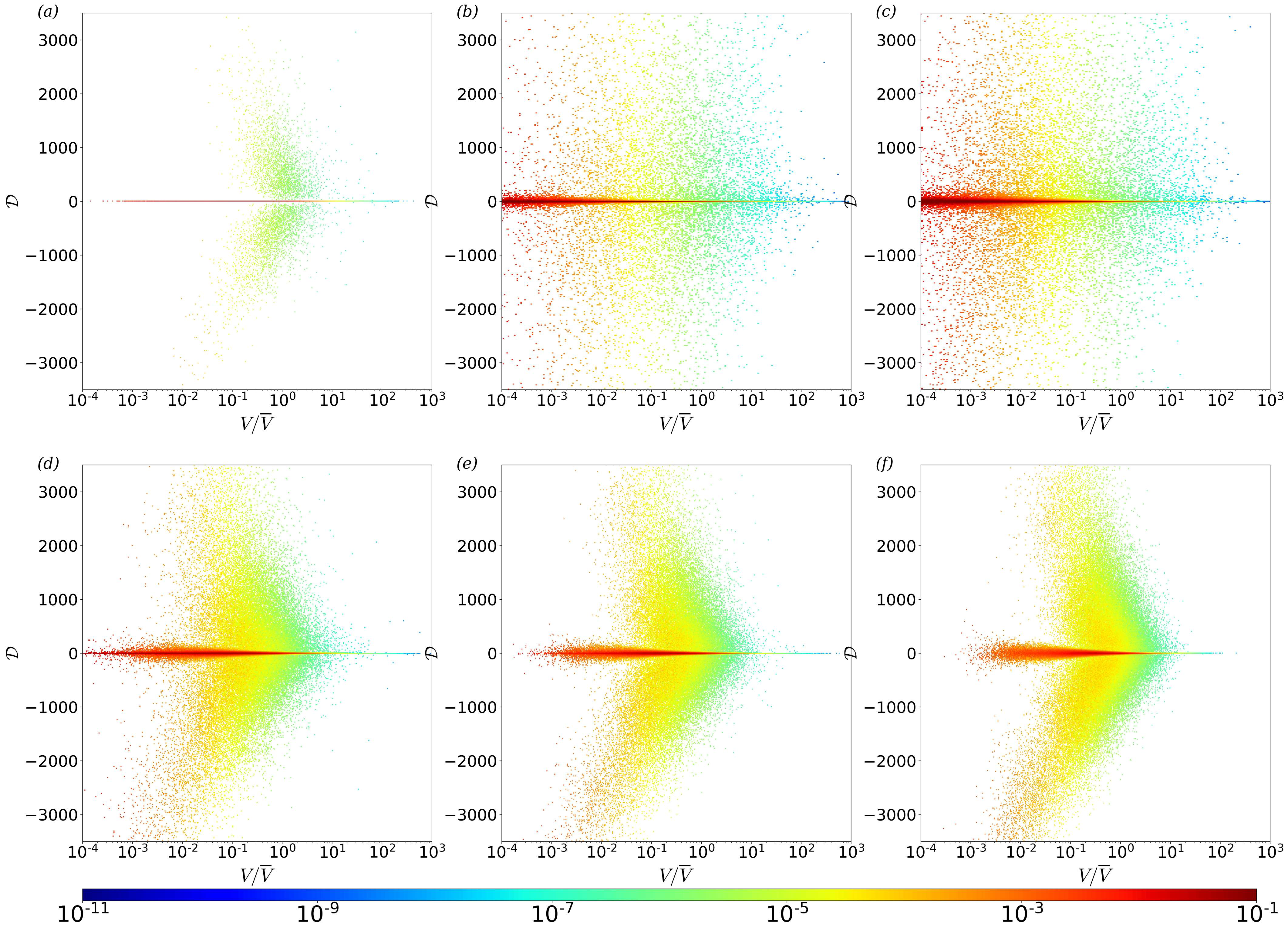}
        \caption[]{Joint PDF of the   volume in log scale and divergence in linear scale for (a) $ St = 0.05$, (b) $0.5$, (c) $1$, (d) $5$, (e) $10$  and (f) $50$ in quasi-adiabatic regime ($c = 0.7$, cHW).}
        \label{fig: joint pdf}
 \end{figure}
 
\begin{figure}
    \centering
    {\includegraphics[width=1.\textwidth]{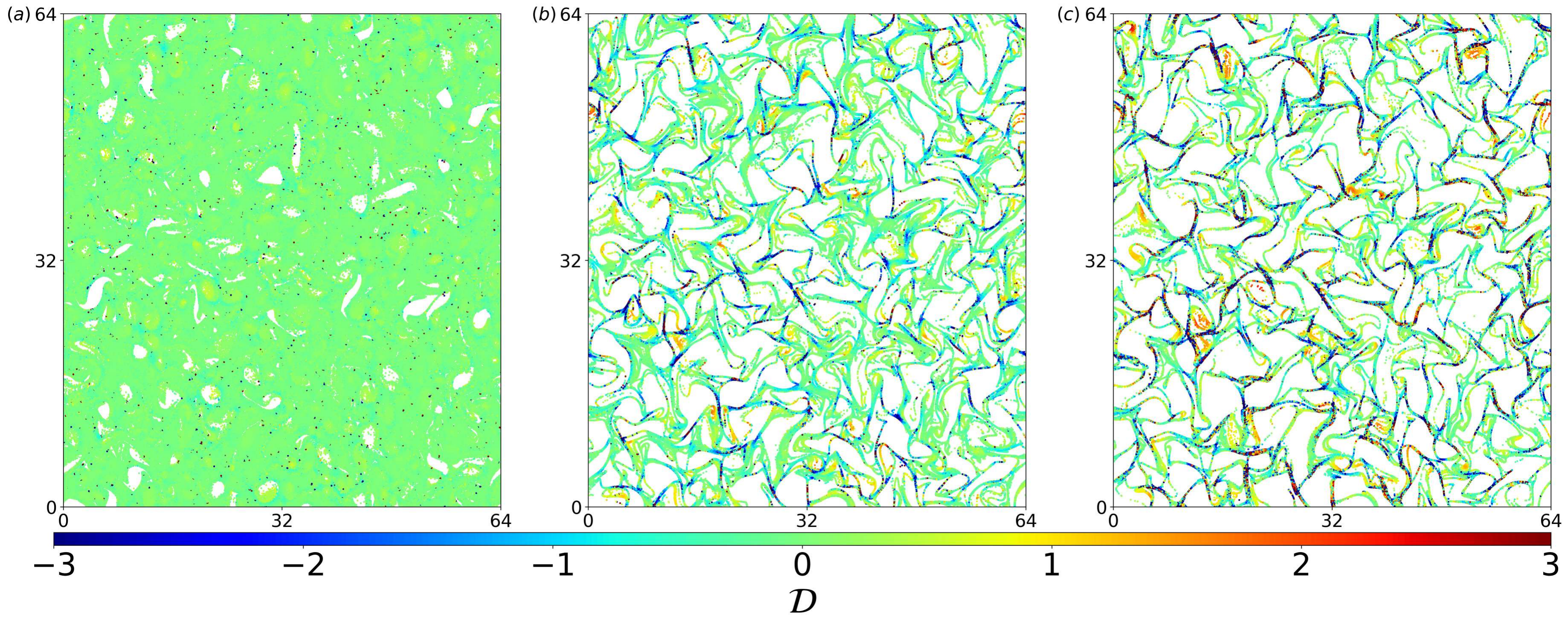}\label{fig: divergence field1}}\\
    {\includegraphics[width=1.\textwidth]{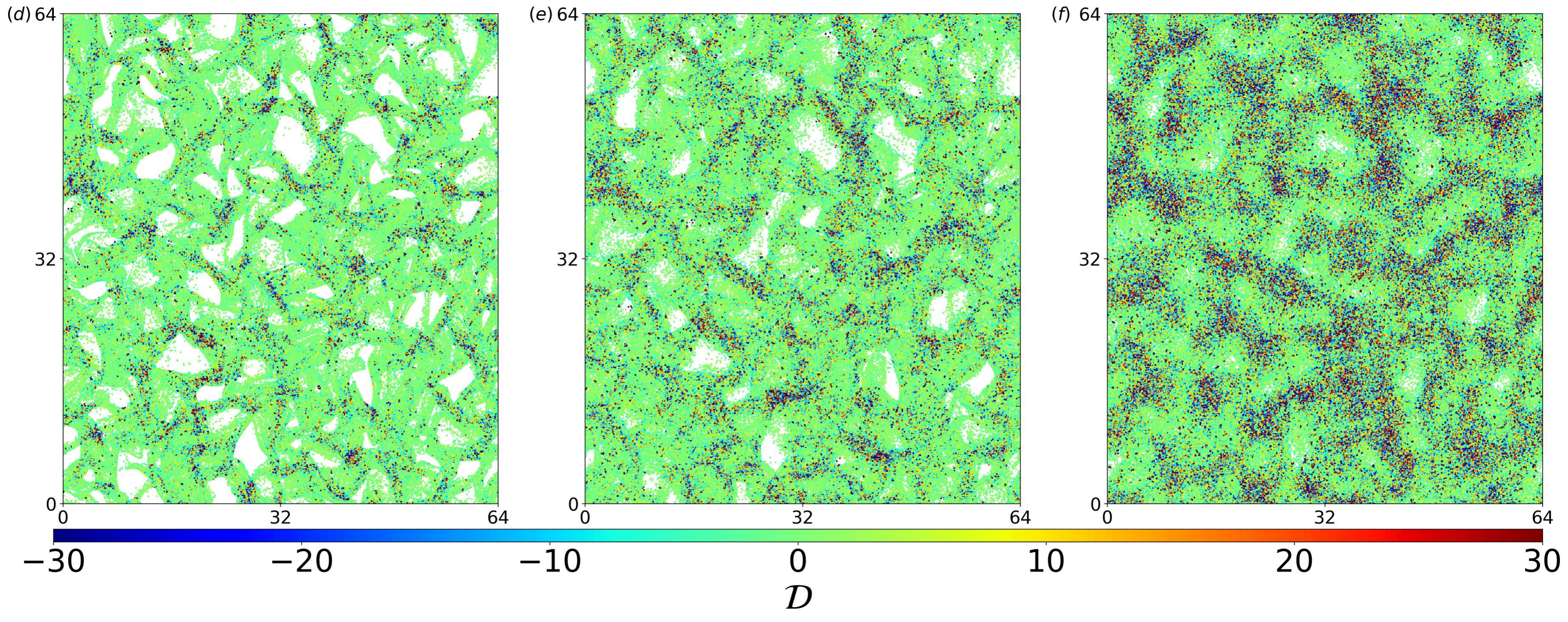}\label{fig: divergence field2}}
    \caption{Spatial distribution of one million particles colored with the divergence $\mathcal{D}$ for(a) $ St = 0.05$, (b) $0.5$, (c) $1$, (d) $5$, (e) $10$  and (f) $50$ in quasi-adiabatic regime ($c = 0.7$, cHW).} 
    \label{fig: divergence field}
\end{figure}

The joint PDFs of the divergence $\mathcal{D}$ and normalized  volume $V/\overline{V}$ show more insight into cluster formation (cf. Figure \ref{fig: joint pdf}). There is a high probability along $\mathcal{D} = 0$, indicating groups of particles move as an incompressible flow. High negative/positive divergence probability in cluster cells ($V/\overline{V}<0.5$) may be due to the fact that the probability of finding particles in clusters is higher than in void region as illustrated by Figure \ref{fig: volume_divergence_pdf}(a). As the Stokes number increases, particles deviate more from the fluid flow, increasing their relative velocity. Therefore, high negative/positive divergence could also be from 'caustics'  where the multi-valued particle velocities at a given position cause large velocity differences between neighboring particles, leading to extreme divergence~\citep{WilkinsonMehlig2005}. Figure \ref{fig: divergence field} visually supports this by coloring particles based on their divergence $\mathcal{D}$. From Figure \ref{fig: divergence field}, we observe that particles with large negative/positive divergence values are densely packed in clusters, indicating simultaneous  particles convergence and divergence.

\section{\label{sec: Conclusions}Conclusions}
In this work, we studied the spatial distribution of  impurity particles in the edge plasma of tokamaks. High-resolution simulations of the plasma flow  were based on the paradigmatic model in edge plasma of tokamaks, namely  Hasegawa--Wakatani model, while the impurity particles were tracked using Lagrangian method. Assuming that impurity particles have a non-zero relaxation time ($\tau_p \neq 0$), which means that  the particles do not immediately adapt to fluctuations in plasma flow, we observed the phenomenon of impurity clustering within the plasma.  To gain deeper insights into the self-organization of impurity particles, we employed a modified Voronoi tessellation technique to compute the divergence of impurity particle velocity $\mathcal{D}$.    The main results are:  
 \begin{enumerate}
    \item \textbf{The impact of $St$  }
    \begin{itemize}
        \item As $St$ increases, impurities cluster in low-vorticity regions due to their inability to adapt to rapid changes in plasma flow. At high $St$, impurities move randomly with less clustering.
        \item With an increase in $St$, there is a higher probability of observing higher values of negative or positive divergence (for $\alpha = 0.22$). This implies that the impurity particle velocity field is converging or diverging faster. These large values of negative or positive divergence are more likely to be found in cluster cells where $(V/\overline{V} < 0.5$).
    \end{itemize}
    \item \textbf{The impact of $\alpha$:}
    \begin{itemize}
        \item The influence of $\alpha$ becomes significant only when $St$ is large, which tends to drive the impurity particles into negative electric potential regions.
    \end{itemize}
\end{enumerate}

For future work, it could be interesting to extend the analysis to multi-resolution techniques that would provide a better understanding of the clustering behavior of impurity particles at different scales,  see e.g. \citet{Matsuda2022}. This approach could reveal more  details about the spatial distribution and dynamics of clusters that are not apparent at a single scale of observation. Another interesting direction is to explore the transport dynamics of heavy impurity particles for different Stokes number by tracking them for a long time. Finally, the limitation of this work is that we  assumed that impurities do not affect plasma flow. While this simplification allowed for a focused study on impurity particle dynamics, it may not capture the full complexity of the interaction between impurities and plasma flow. A possible extension in future studies is to explore the impact of impurity particles on the plasma flow  based on the work of~\citet{benkadda1996nonlinearities}.

\section*{Acknowledgements}

ZL acknowledges  Philipp Krah for the insightful discussions. Centre de Calcul Intensif d’Aix-Marseille is acknowledged for providing access to its high performance computing resources.

\section*{Funding}
This work was supported by I2M (Z.L., T.M.O., K.S.); the French Federation for Magnetic Fusion Studies (FR-FCM) and the Eurofusion consortium, funded by the Euratom Research and Training Programme (Z.L., T.M.O., K.S., S.B., grant number 633053); and the Agence Nationale de la Recherche (ANR), project CM2E (Z.L., T.M.O., B.K., K.S., grant number ANR-20-CE46-0010-01). The views and opinions expressed herein do not necessarily reflect those of the European Commission.
 
\section*{Declaration of interests}
The author reports no conflict of interest.
\appendix

\section{\label{appendix: different regimes}The behavior of $_{74}^{184}\mathrm{W}^{20+}$  in different flow regimes}
\subsection{Hydrodynamic and adiabatic regime}
\begin{figure}
        \centering
        \includegraphics[width=0.9\linewidth]{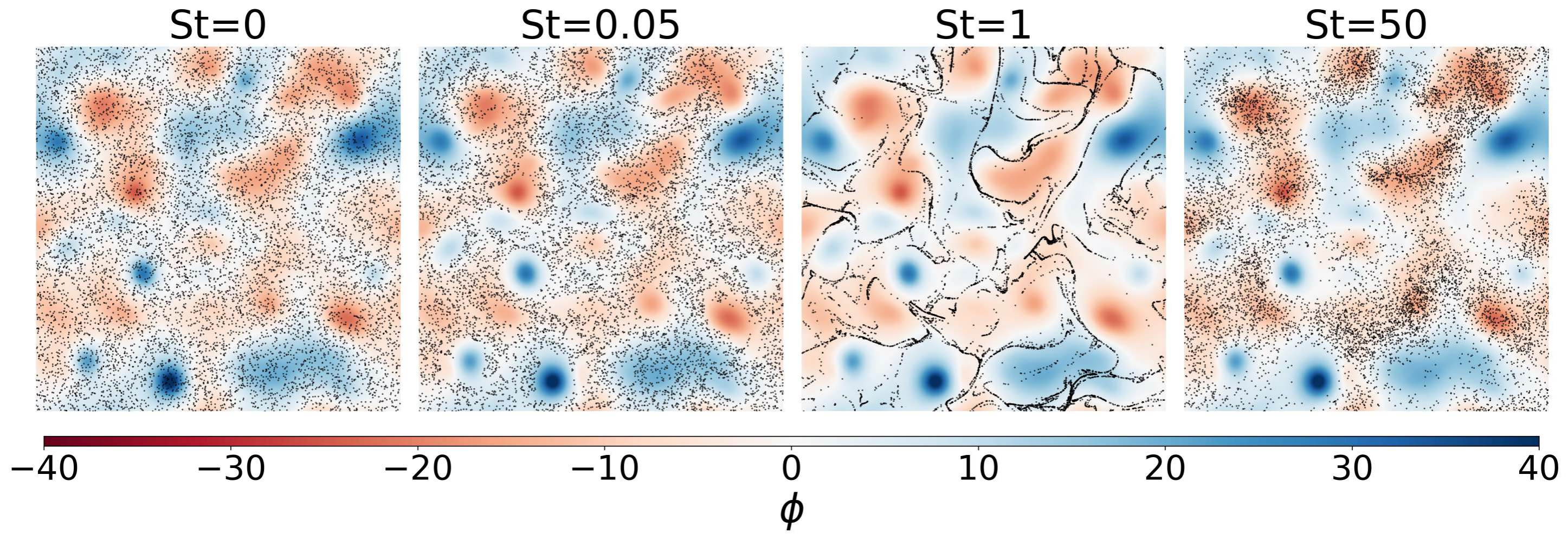}
        \caption[]{Electric potential fields $\phi$ (stream function) superimposing $10^4$ impurity particles (out of
$10^6$) for various Stokes numbers in statistically steady state within the hydrodynamic regime ($c = 0.01$, cHW).}
        \label{fig: vorticity and particles: c =0.01, cHW}
 \end{figure}
\begin{figure}
        \centering
        \includegraphics[width=0.9\linewidth]{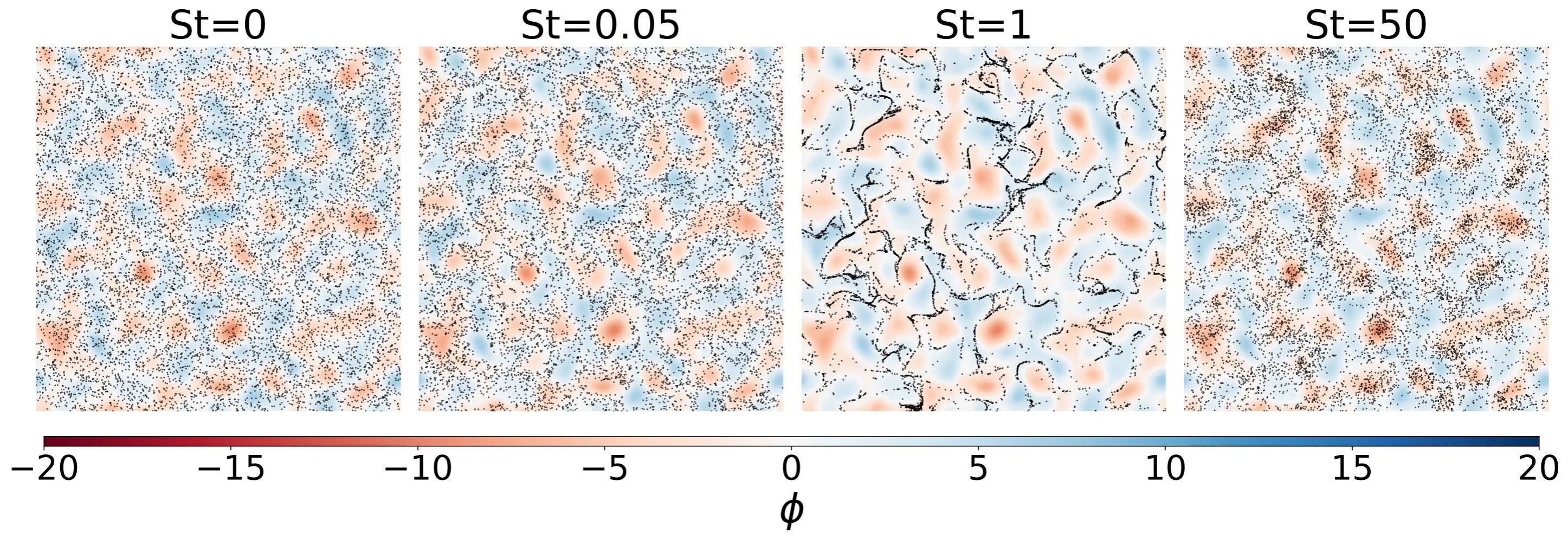}
        \caption[]{Electric potential fields $\phi$ (stream function) superimposing $10^4$ impurity particles (out of $10^6$) for various Stokes numbers in statistically steady state within the adiabatic regime ($c = 2$, cHW).}
        \label{fig: vorticity and particles: c = 2, cHW}
 \end{figure}

\begin{figure}
    \centering
    \subfloat[]{\includegraphics[width=0.48\textwidth]{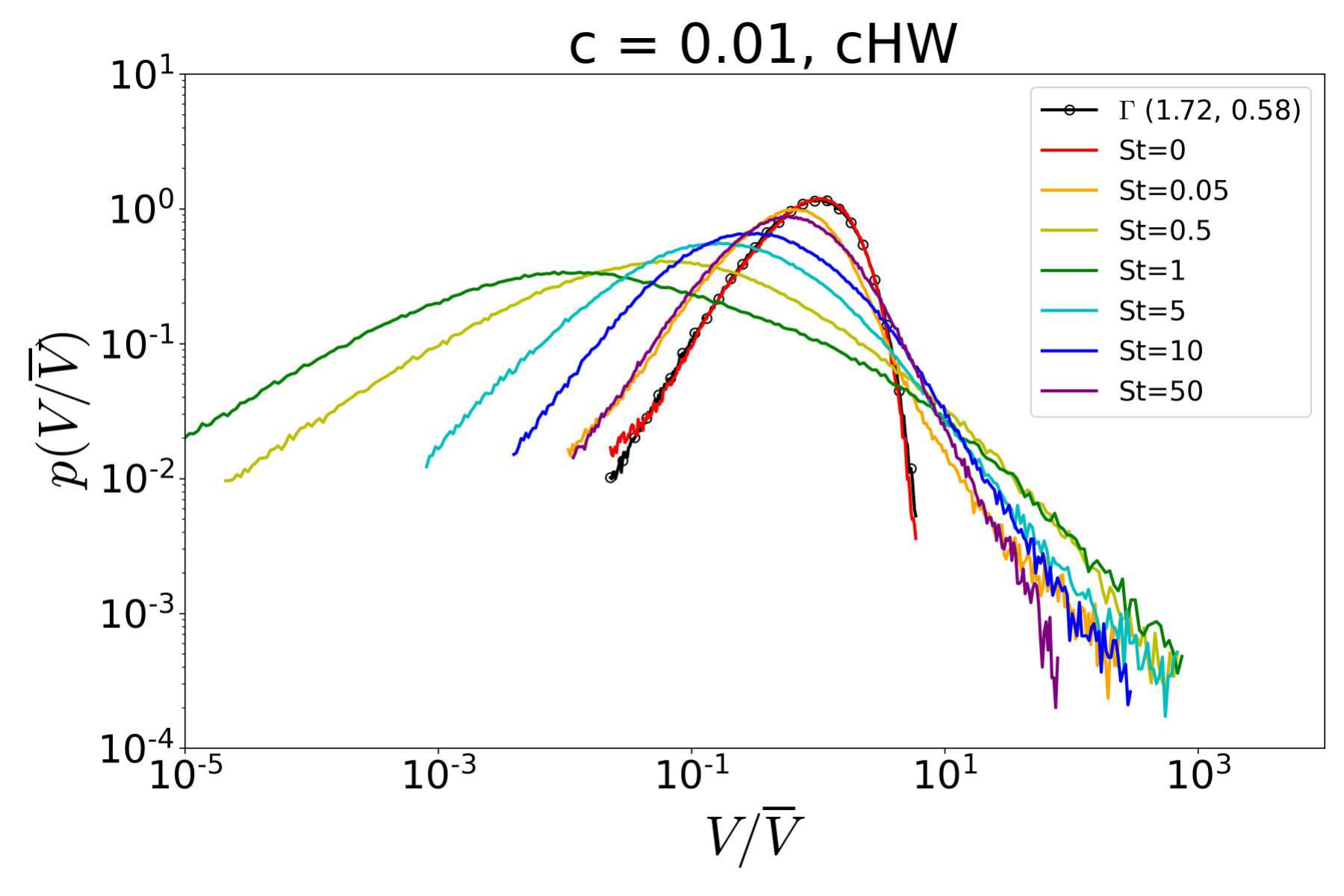}}\quad
    \subfloat[]{\includegraphics[width=0.48\textwidth]{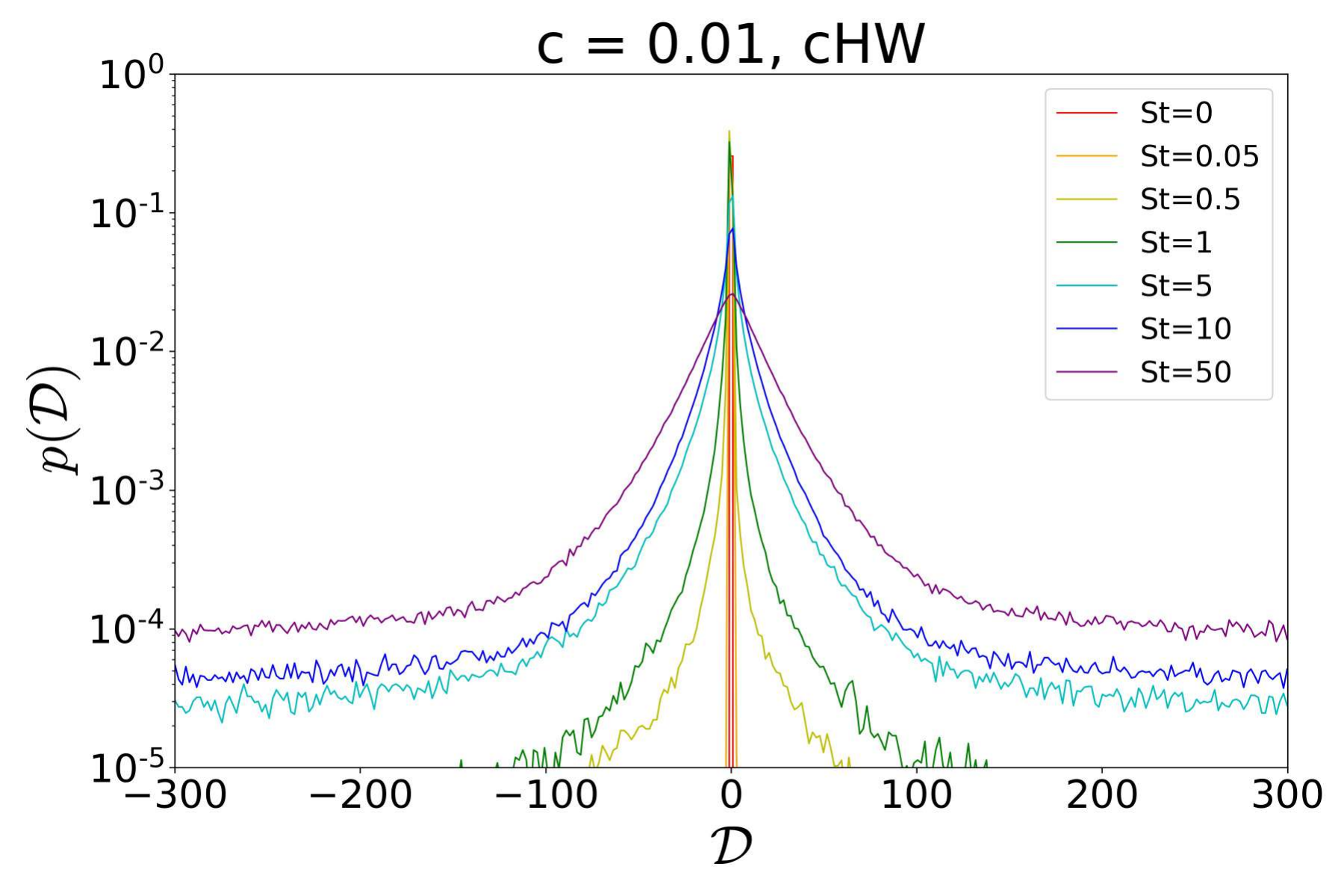}}\label{fig: divergence_pdf}
    \caption{(a) PDF of  volume normalized by the mean ($V/\overline{V}$) (b) PDF of impurity velocity divergence $\mathcal{D}$ for different Stokes numbers for $_{74}^{184}\mathrm{W}^{20+}(\alpha = 0.22)$ in hydrodynamic regime ($c = 0.01$, cHW)} 
    \label{fig: volume_divergence_pdf_C0_01}
\end{figure}

\begin{figure}
    \centering
    \subfloat[]{\includegraphics[width=0.48\textwidth]{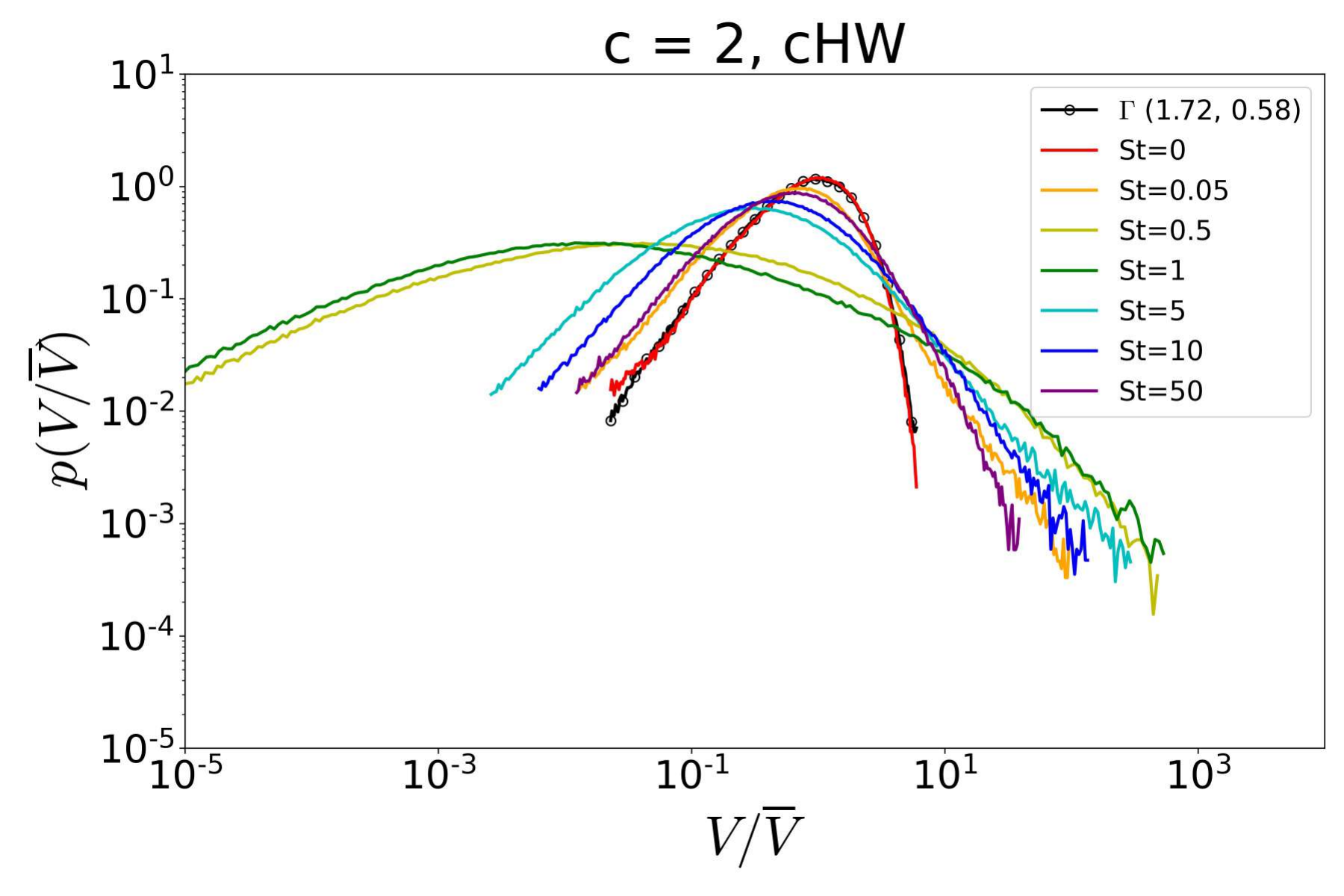}}\quad
    \subfloat[]{\includegraphics[width=0.48\textwidth]{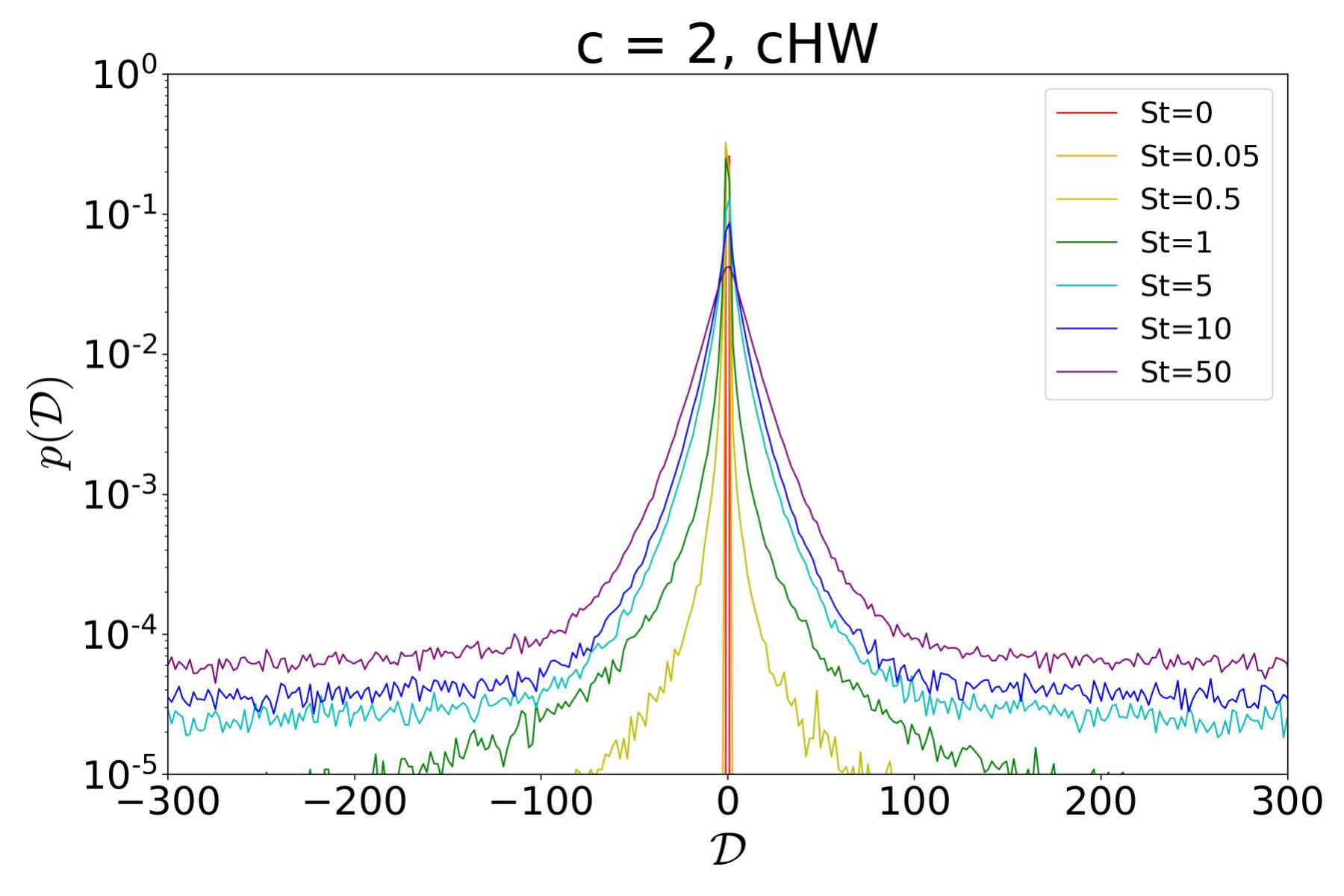}}\label{fig: divergence_pdf}
    \caption{(a) PDF of  volume normalized by the mean ($V/\overline{V}$) (b) PDF of impurity velocity divergence $\mathcal{D}$ for different Stokes numbers for $_{74}^{184}\mathrm{W}^{20+}(\alpha = 0.22)$ in hydrodynamic regime ($c = 2$, cHW)} 
    \label{fig: volume_divergence_pdf_C2}
\end{figure}
We explored the  behavior of $_{74}^{184}\mathrm{W}^{20+}$  in hydrodynamic ($c = 0.01$, cHW) and adiabatic regime ($c = 2$, cHW). Figures \ref{fig: vorticity and particles: c =0.01, cHW} to \ref{fig: vorticity and particles: c = 2, cHW} show potential fields and impurity behavior in hydrodynamic  and adiabatic   regimes. We have the same observation as quasi-adiabatic case ($c = 0.7$, cHW): at $St = 0$, impurities act as passive tracers, uniformly distributed along fluid streamlines; as $St= 0.05$, particles start to cluster in areas of low vorticity and  clustering is clear at $St = 1$ ;  at $St = 50$,  there is less clustering and  particles are mostly presented in negative potential areas. Similar to the quasi-adiabatic case,   As shown in Figure~\ref{fig: volume_divergence_pdf_C0_01}(a) and~\ref{fig: volume_divergence_pdf_C2}(a), as $St$ increases, the number of cluster cells ($V/\overline{V} < 0.5$) initially increases until $St = 1$. Figure~\ref{fig: volume_divergence_pdf_C0_01}(b) and~\ref{fig: volume_divergence_pdf_C2}(b) demonstrate that the PDF of divergence $\mathcal{D}$ broadens with increasing Stokes number.

\subsection{Zonal flows}
\begin{figure}
        \centering
        \includegraphics[width=0.9\linewidth]{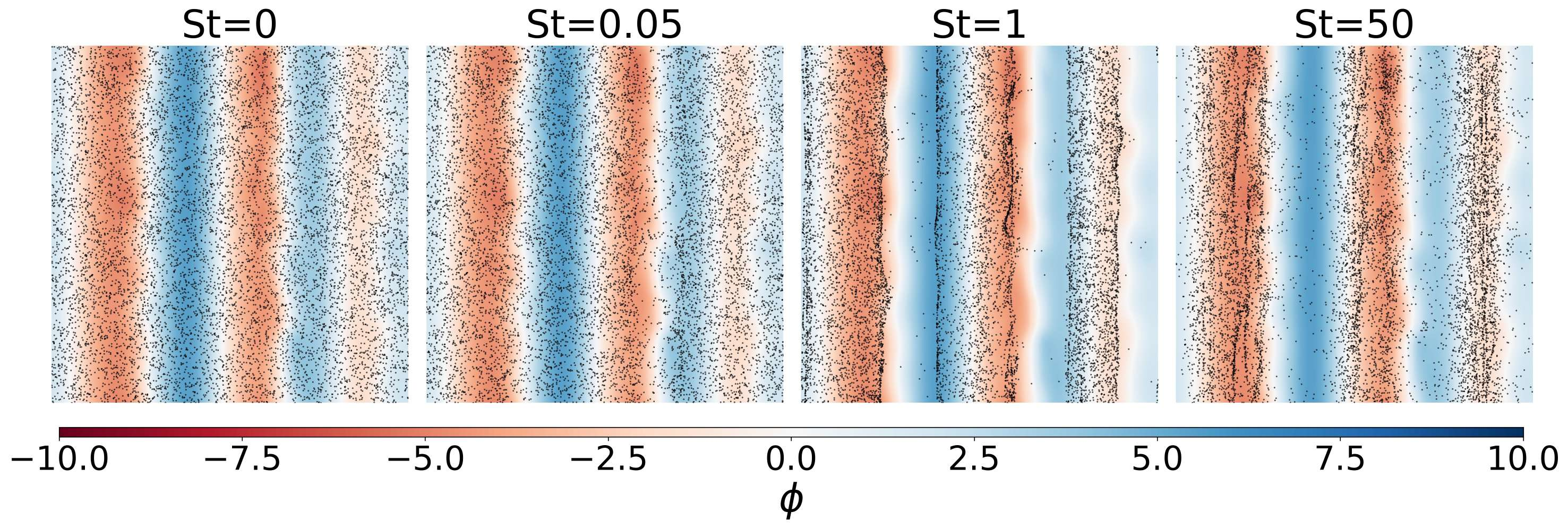}
        \caption[]{Electric potential fields $\phi$   superimposing $10^4$ impurity particles (out of $10^6$) for various Stokes numbers in statistically steady state within zonal flows ($c = 2$, mHW).}
        \label{fig: potential and particles: c = 2, mHW}
 \end{figure}

\begin{figure}
        \centering
        \includegraphics[width=0.4\linewidth]{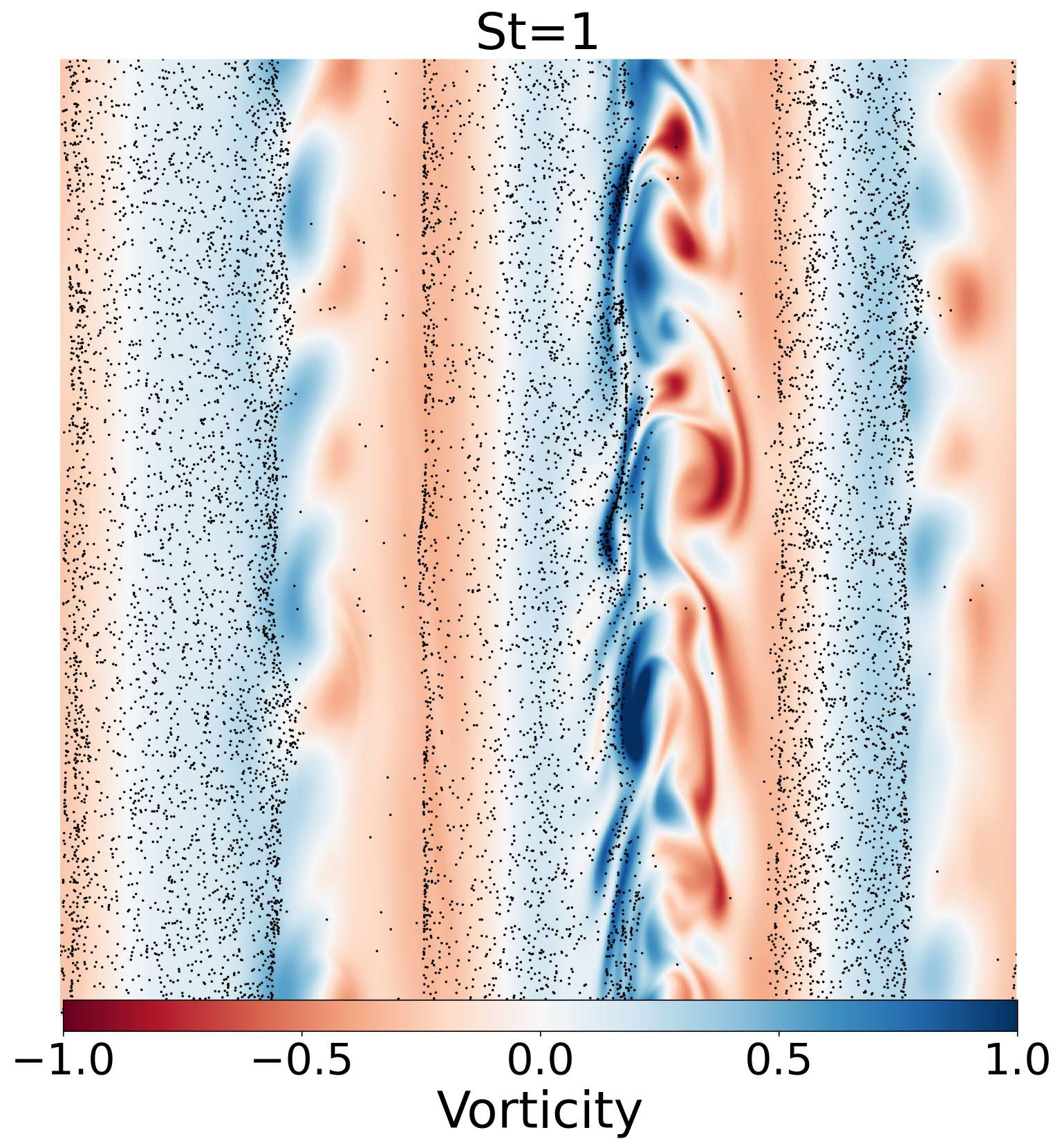}
        \caption[]{Vorticity field superimposing $10^4$ impurity particles (out of $10^6$) for various Stokes numbers in statistically steady state within zonal flows ($c = 2$, mHW).}
        \label{fig: vorticity and particles: c = 2, mHW}
 \end{figure}

 \begin{figure}
    \centering
    \subfloat[]{\includegraphics[width=0.48\textwidth]{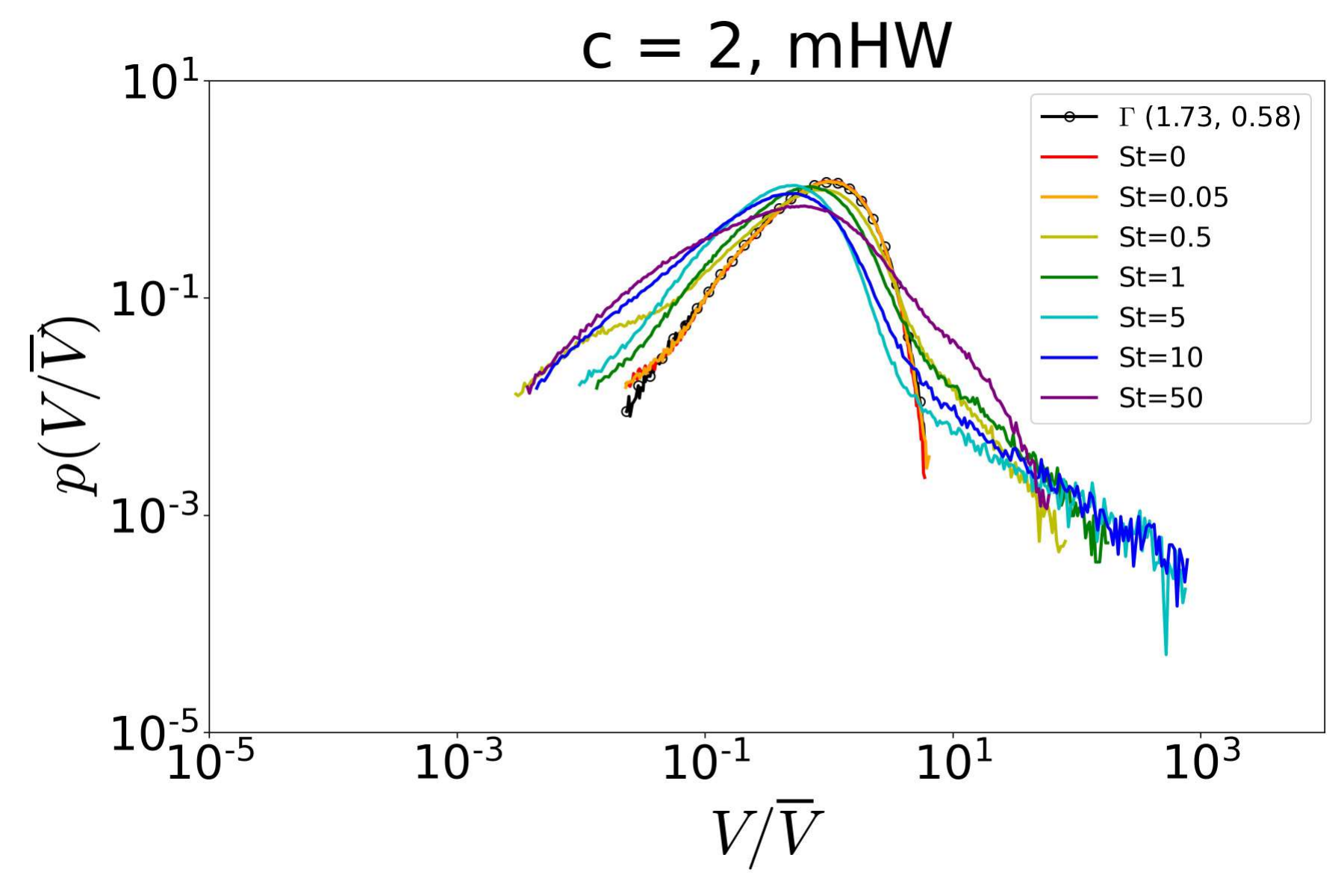}}\quad
    \subfloat[]{\includegraphics[width=0.48\textwidth]{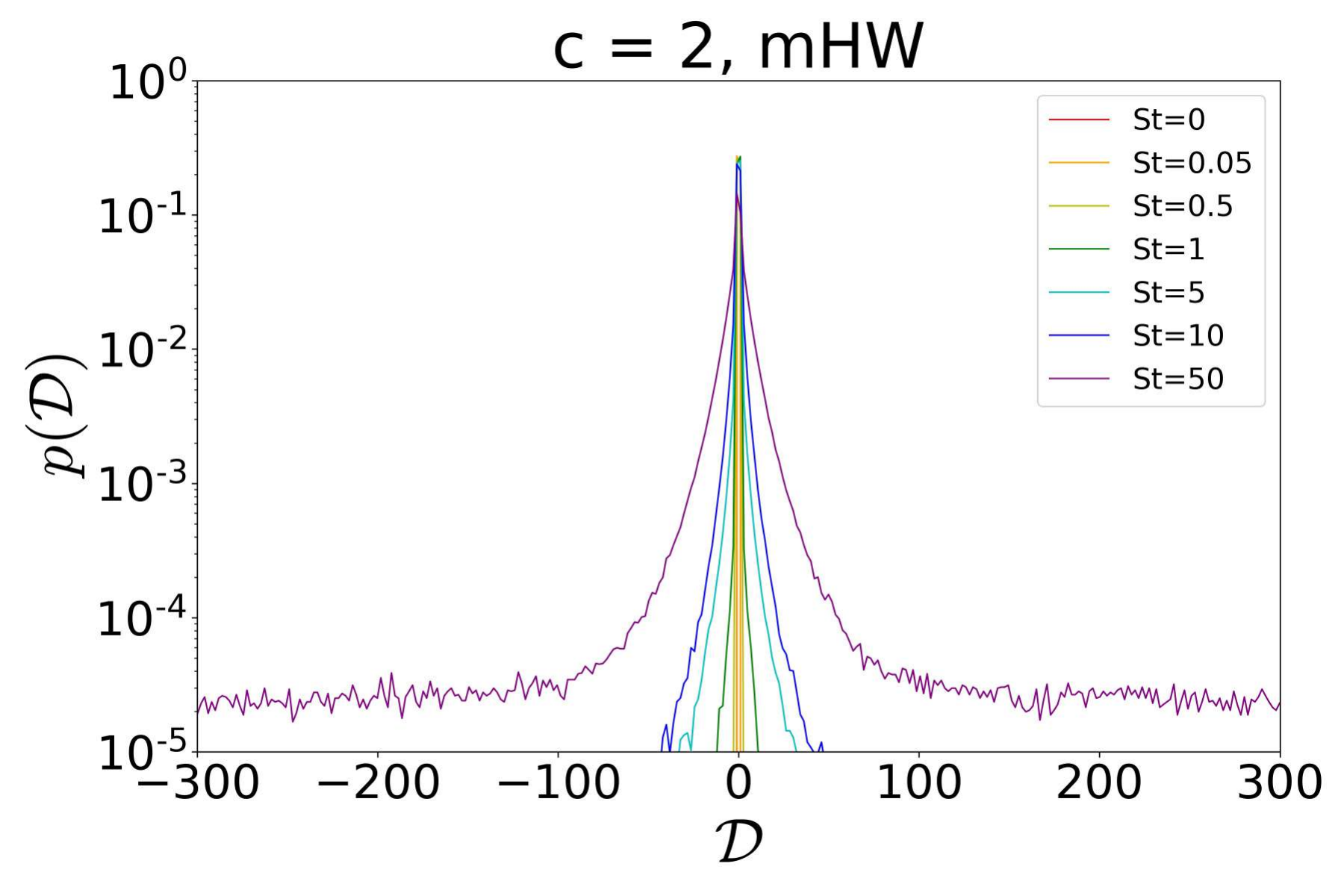}}\label{fig: divergence_pdf}
    \caption{(a) PDF of  volume  normalized by the mean ($V/\overline{V}$)   (b) PDF of the impurity velocity divergence $\mathcal{D}$ for different Stokes numbers   for  $_{74}^{184}\mathrm{W}^{20+}(\alpha = 0.22)$ in zonal flows ($c = 2$, mHW).} 
    \label{fig: volume_divergence_pdf_a_0_22_c2_mHW}
\end{figure}

In the modified Hasegawa-Wakatani model (mHW) with  $c = 2$, zonal flows are present. Figure~\ref{fig: potential and particles: c = 2, mHW} shows the zonal flows and superimposing impurity particles. While it may not be clearly visible that particles with  $St =  1$ are primarily presented in regions of low vorticity, as indicated by Figure~\ref{fig: potential and particles: c = 2, mHW}, it is evident in the vorticity field (Figure~\ref{fig: vorticity and particles: c = 2, mHW}). For  $St= 50$, particles are predominantly found in regions of negative potential. As shown in Figure~\ref{fig: volume_divergence_pdf_a_0_22_c2_mHW}(a), there are fewer cluster cells ($V/\overline{V} < 0.5$) compared to other regimes where  zonal flows are absent ($c = 0.01, 0.7 ,2$, cHW). Figure~\ref{fig: volume_divergence_pdf_a_0_22_c2_mHW}(b) illustrates that, in the presence of zonal flows, the likelihood of large negative or positive divergence is reduced compared to other regimes, suggesting that zonal flows tend to prevent the convergence and divergence of particles.

\section{\label{appendix: Analysis}Analysis of Tungsten impurity with different charge state in quasi-adiabatic regime ($c = 0.7$, cHW)}
 
\begin{figure}
    \centering
    \subfloat[]{\includegraphics[width=0.48\textwidth]{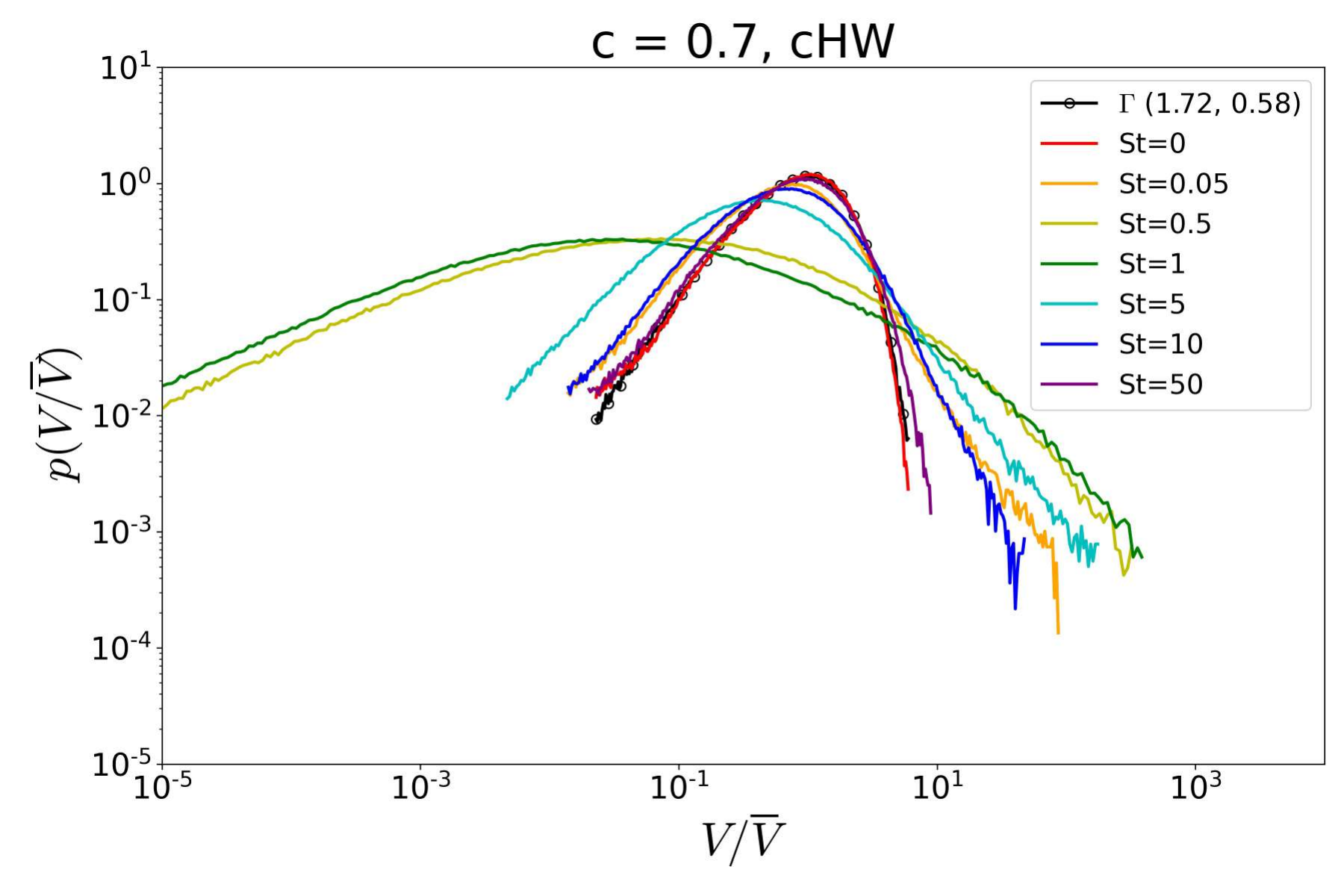}}\quad
    \subfloat[]{\includegraphics[width=0.48\textwidth]{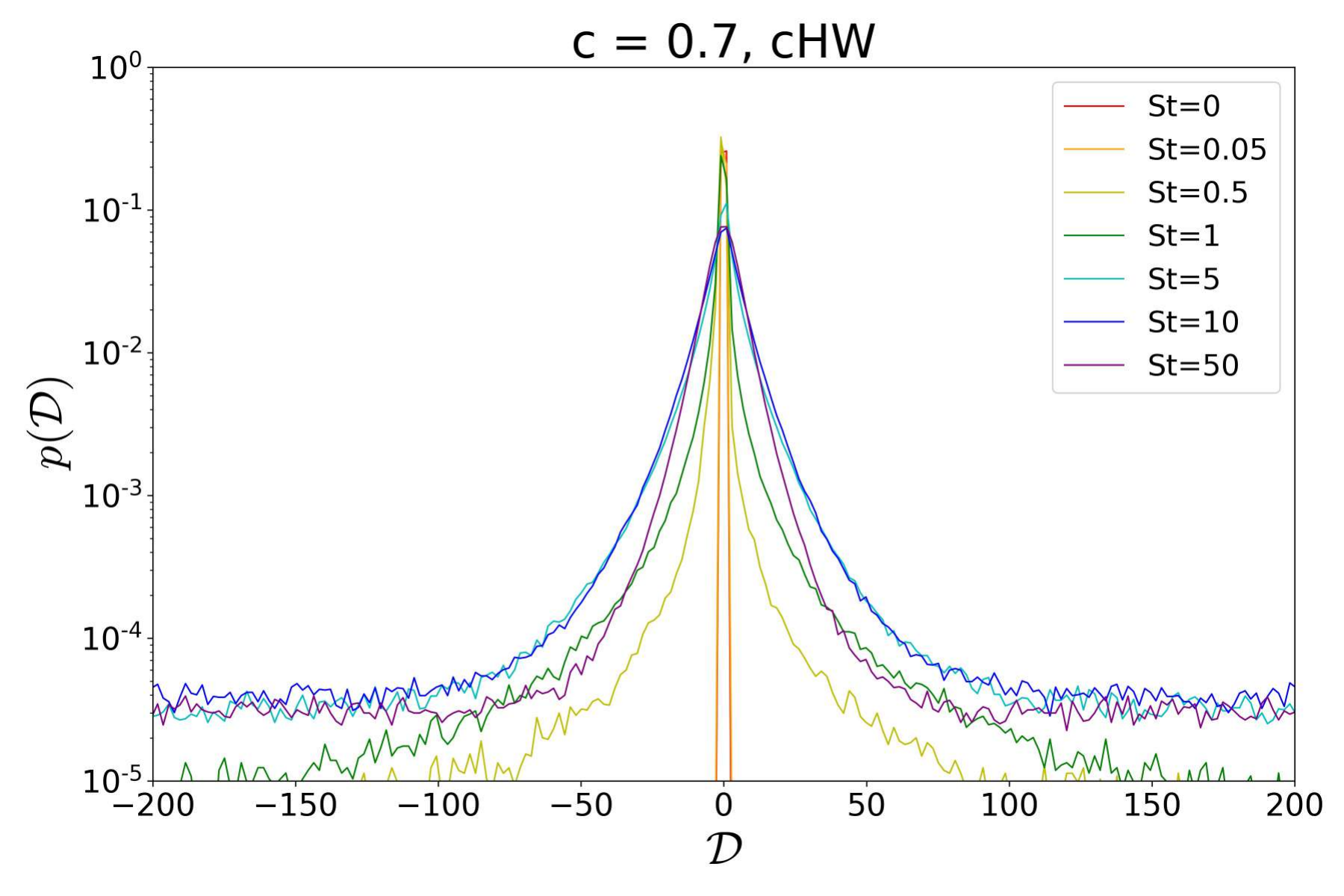}}\label{fig: divergence_pdf}
    \caption{(a) PDF of  volume normalized by the mean ($V/\overline{V}$) (b) PDF of impurity velocity divergence $\mathcal{D}$ for different Stokes numbers   for  $_{74}^{184}\mathrm{W}^{1+}(\alpha = 0.01)$ in quasi-adiabatic regime ($c = 0.7$, cHW).} 
    \label{fig: volume_divergence_pdf_a_0_01}
\end{figure}

\begin{figure}
    \centering
    \subfloat[]{\includegraphics[width=0.48\textwidth]{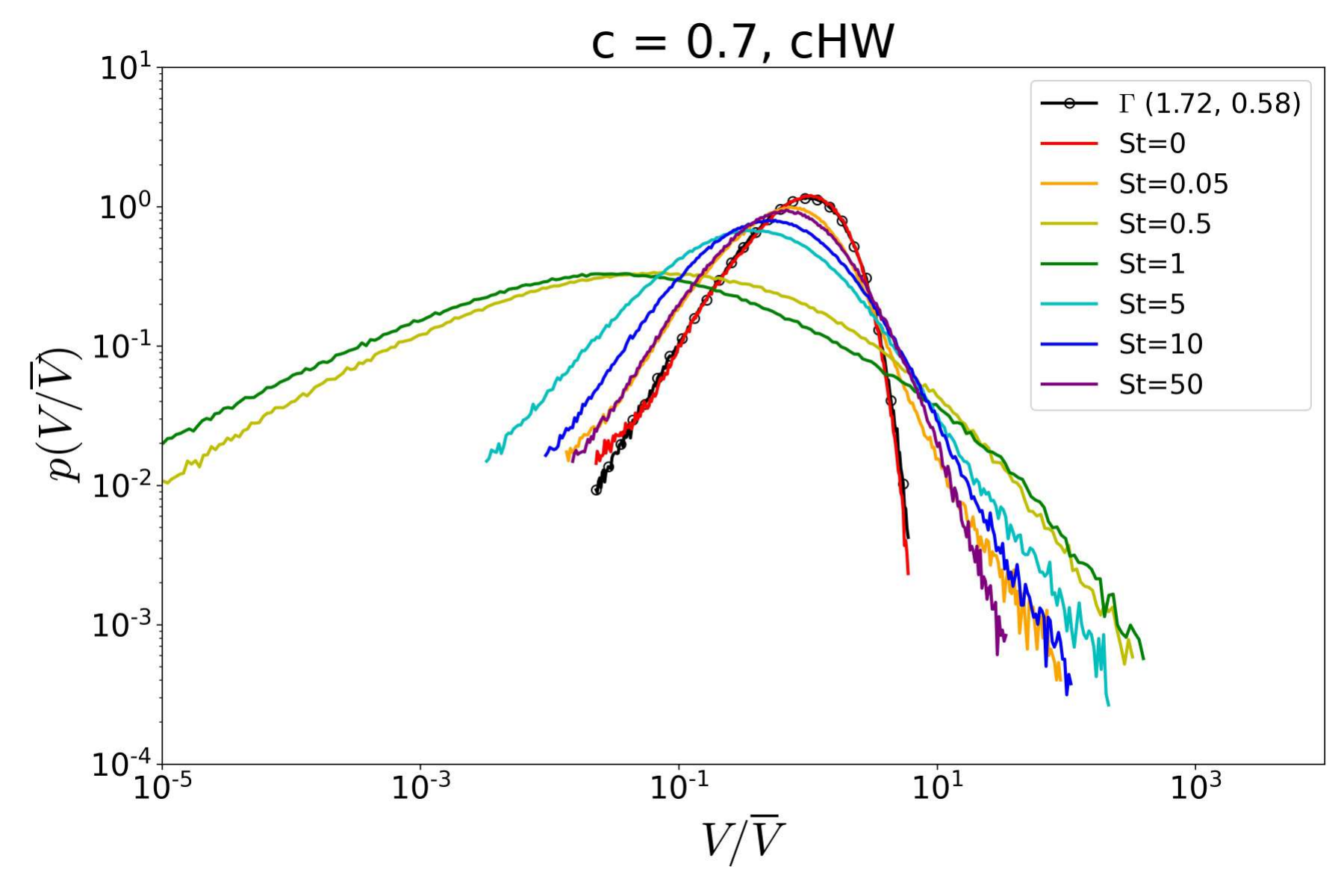}}\quad
    \subfloat[]{\includegraphics[width=0.48\textwidth]{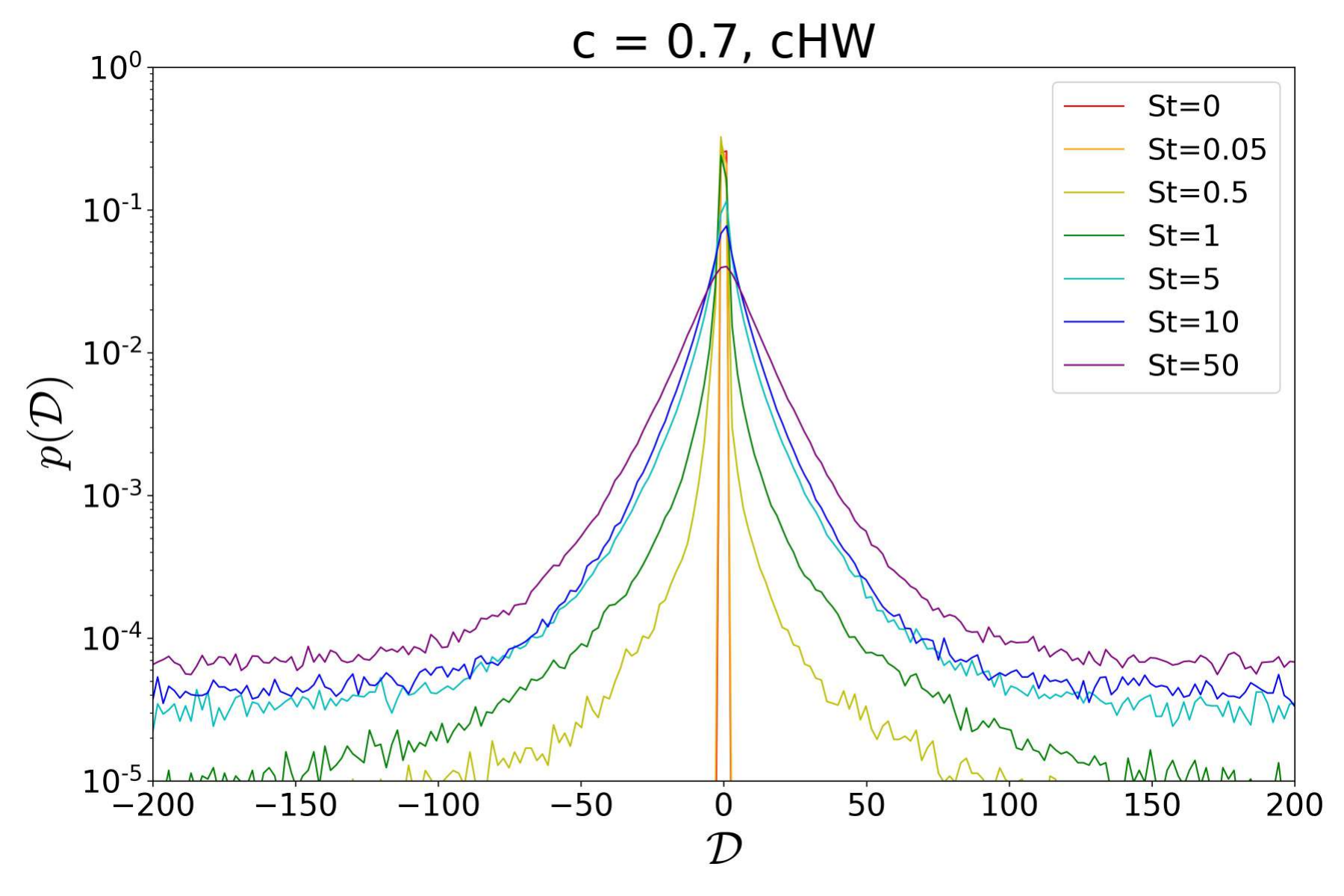}}\label{fig: divergence_pdf}
    \caption{(a) PDF of  volume  normalized by the mean ($V/\overline{V}$)   (b) PDF of the impurity velocity divergence $\mathcal{D}$ for different Stokes numbers   for  $_{74}^{184}\mathrm{W}^{44+}(\alpha = 0.48)$ in quasi-adiabatic regime ($c = 0.7$, cHW).} 
    \label{fig: volume_divergence_pdf_a_0_48}
\end{figure}

The clustering of two other Tungsten impurities with different charge state is analyzed in this appendix. We consider $_{74}^{184}\mathrm{W}^{1+}$ ($\alpha = 0.01$) and $_{74}^{184}\mathrm{W}^{44+}(\alpha = 0.48)$ \citep{maget2020natural}. Applying modified Voronoi tesselation, the PDF of   volume normalized by the mean ($V/\overline{V}$) and the PDF of impurity velocity divergence $\mathcal{D}$ for different Stokes numbers are plotted in Figure~\ref{fig: volume_divergence_pdf_a_0_01} and Figure~\ref{fig: volume_divergence_pdf_a_0_48}.

Figure~\ref{fig: volume_divergence_pdf_a_0_01}(a) and Figure~\ref{fig: volume_divergence_pdf_a_0_48}(a) show that for both $_{74}^{184}\mathrm{W}^{1+}$  and $_{74}^{184}\mathrm{W}^{44+}$, at $St = 0$, the impurity particles exhibit a random distribution, as evidenced by the alignment of the curve with the gamma distribution.  This indicates the absence of   inertial effects. As $St$ increases, the number of cluster cells ($V/\overline{V} < 0.5$) increases and then decreases after exceeding $St =1$.  From Figure~\ref{fig: volume_divergence_pdf_a_0_01}(b), for $_{74}^{184}\mathrm{W}^{1+}$, it is observed that the PDFs of divergence broaden with increasing $St$,  then the PDFs begin to narrow from $St = 10$ to $St = 50$.  From Figure~\ref{fig: volume_divergence_pdf_a_0_01}(b), for $_{74}^{184}\mathrm{W}^{ 44+}$, it is observed  that the PDFs of divergence widen as $St$ increases.

Tables~\ref{tab: mean of positive divergence} and~\ref{tab: variance of divergence} provide a quantitative comparison of how the divergence characteristics change with varying $\alpha$ values, which are indicative of the charge state of the Tungsten ions ($\alpha = {Zm_i}/{m_{p}}$). Given that negative and positive divergence values are  comparable, the overall average divergence approaches zero. So we calculate the mean of positive divergence $\overline{\mathcal{D}_+}$, which, by implication, mirrors the behavior observed for negative divergence. For low Stokes numbers ($St = 0.05, 0.5, 1$), the mean of positive divergence ($\overline{\mathcal{D}_+}$) remains relatively consistent regardless of the ionization state, suggesting that the charge state has a minimal impact on the divergence behavior of the impurities in this range. However, as we examine higher Stokes numbers ($St = 5, 10, 50$), a clear trend emerges: both the variance of divergence ($\sigma^2_{\mathcal{D}}$) and the mean of positive divergence ($\overline{\mathcal{D}_+}$) increase with higher $\alpha$ values. This observation implies that at higher $St$ values, the divergence behavior of the impurities is  influenced by their charge state.

\begin{table}
\begin{center}
\setlength{\tabcolsep}{8pt} 
\begin{tabular}{cccccccc}
$St$  & 0.05 & 0.5 & 1 & 5 & 10 & 50 \\

$\overline{\mathcal{D}_+}(\alpha = 0.01)$   & 0.06 & 0.4 & 1.5 & 9.2 & 15.1  & 9.0  \\
$\overline{\mathcal{D}_+}(\alpha = 0.22)$   & 0.06 & 0.4 & 1.5 & 9.5 & 17.4  & 39.8  \\
$\overline{\mathcal{D}_+}(\alpha = 0.48)$   & 0.06 & 0.4 & 1.4 & 10.9 & 24.69 & 63.0

\end{tabular}
\caption{Mean of positive divergence $\overline{\mathcal{D}_+}$  for different Stokes numbers and different $\alpha$ values.}
\label{tab: mean of positive divergence}
\end{center}
\end{table}

\begin{table}
\begin{center}
\setlength{\tabcolsep}{8pt} 
\begin{tabular}{cccccccc}
$St$   & 0.05 & 0.5 & 1 & 5 & 10 & 50 \\
$\sigma^2_{\mathcal{D}}$($\alpha = 0.01$)  & 0.009 & 1.1 & 10.7  & 1106  & 3153  & 1087  \\
$\sigma^2_{\mathcal{D}}$($\alpha = 0.22$)  & 0.009 & 1.0  & 10.6  & 1214  & 4126  & 15902  \\
$\sigma^2_{\mathcal{D}}$($\alpha = 0.48$)   & 0.009 &1.0 & 9.3  & 1632 & 7529  & 31319 
\end{tabular}
\caption{Variance of divergence $\sigma^2_{\mathcal{D}}$ for different Stokes numbers and different $\alpha$ values.}
\label{tab: variance of divergence}
\end{center}
\end{table}

\newpage
\bibliographystyle{jpp}

\bibliography{jpp-instructions}

\end{document}